\documentclass[journal, 11pt]{IEEEtran}
\IEEEoverridecommandlockouts
\usepackage{graphicx}
\usepackage{caption}
\usepackage{subcaption}
\usepackage{mathtools}
\usepackage{bm,amsmath, amssymb}
\usepackage{amsmath}
\usepackage{amsfonts}
\usepackage {amssymb,amsmath}
\usepackage{cite}
\usepackage{epstopdf}
\usepackage{color}
\usepackage{microtype}
\DeclareGraphicsExtensions{.pdf,.png,.jpg}

\newcommand {\Define} {\stackrel {\Delta} {=}  }

\usepackage{cite}
\usepackage[margin=2cm]{geometry}
\usepackage[most]{tcolorbox}
\usepackage[normalem]{ulem}

\newtcolorbox{mytextbox}[1][]{%
  sharp corners,
  enhanced,
  colback=white,
  attach title to upper,
  #1
}

\def\BibTeX{{\rm B\kern-.05em{\sc i\kern-.025em b}\kern-.08em
    T\kern-.1667em\lower.7ex\hbox{E}\kern-.125emX}}
    
\begin{document}
\title{{Does 6G Need a New Waveform: Comparing Zak-OTFS with CP-OFDM}}
\author{\IEEEauthorblockN{Imran Ali Khan\IEEEauthorrefmark{1}, Saif Khan Mohammed\IEEEauthorrefmark{1}, 
Ronny Hadani\IEEEauthorrefmark{2}\IEEEauthorrefmark{3},
Ananthanarayanan Chockalingam\IEEEauthorrefmark{4}, \\
Robert Calderbank\IEEEauthorrefmark{5},~\IEEEmembership{Fellow,~IEEE},  Anton Monk\IEEEauthorrefmark{3}, Shachar Kons\IEEEauthorrefmark{3}, Shlomo Rakib\IEEEauthorrefmark{3},
and Yoav Hebron\IEEEauthorrefmark{3}
\thanks{This work has been submitted to the IEEE for possible publication. Copyright may be transferred without notice, after which this version may no longer be accessible.}
}

\vspace{5mm}

\IEEEauthorblockA{\IEEEauthorrefmark{1}Department of Electrical Engineering, Indian Institute of Technology Delhi, India}\\
\IEEEauthorblockA{\IEEEauthorrefmark{2}Department of Mathematics, University of Texas at Austin, USA}\\
\IEEEauthorblockA{\IEEEauthorrefmark{3} Cohere Technologies Inc., San Jose, CA, USA}\\
\IEEEauthorblockA{\IEEEauthorrefmark{4} Department of Electrical Communication Engineering, Indian Institute of Science Bangalore, India}
\IEEEauthorblockA{\IEEEauthorrefmark{5}Department of Electrical and Computer Engineering, Duke University, USA}}

\maketitle

\vspace{-6mm}
\begin{abstract}
Across the world, there is growing interest in new waveforms, Zak-OTFS in particular, and over-the-air implementations are starting to appear. The choice between OFDM and Zak-OTFS is not so much a choice between waveforms as it is an architectural choice between preventing inter-carrier interference (ICI) and embracing ICI. In OFDM, once the Input-Output (I/O) relation is known, equalization is relatively simple, at least when there is no ICI. However, in the presence of ICI the I/O relation is non-predictable and its acquisition is non-trivial. In contrast, equalization is more involved in Zak-OTFS due to inter-symbol-interference (ISI), however the I/O relation is predictable and its acquisition is simple. {Zak-OTFS exhibits superior performance in doubly-spread 6G use cases with high delay/Doppler channel spreads (i.e., high mobility and/or large cells), but architectural choice is governed by the typical use case, today and in the future. What is typical depends to some degree on geography, since large delay spread is a characteristic of large cells which are the rule rather than the exception in many important wireless markets.} This paper provides a comprehensive performance comparison of cyclic prefix OFDM (CP-OFDM) and Zak-OTFS across the full range of 6G propagation environments. The performance results provide insights into the fundamental architectural choice.
\end{abstract}

\section{Introduction}
Standardization for sixth generation (6G) communication technologies based on IMT-2030 is already underway \cite{IMT2030, 3gpptimeline}. New 6G use cases, for example Integrated Sensing and Communication (ISAC), ubiquitous communication based on Non-terrestrial networks (Leo-satellite communication), AI and ML in wireless communication, bring new challenges which has initiated industry-wide discussions on waveforms for 6G \cite{ieee6gpaper}. While Orthogonal Frequency Division Multiplexing (OFDM) has served well in 4G/5G, it might not be suitable for all 6G use cases, and therefore a growing interest in new waveforms \cite{Shafi2025,airtoground, LMLC}. Zak-Orthogonal Time Frequency Space (Zak-OTFS) is a new waveform which is a leading candidate waveform for high Doppler (high mobility) and delay spread use cases in 6G, due to its robustness to mobility induced Doppler shifts and high channel delay spreads \cite{zakotfs1, zakotfs2, otfsbook}.

We provide the first comprehensive comparison of the spectral efficiencies of CP-OFDM and Zak-OTFS across a full range of 6G propagation environments. We aim to derive fundamental understanding of how relative spectral efficiency (SE) depends on mobility (high/low) and on cell size (large/small). We present the choice between waveforms, as an architectural choice between acquiring interference (Zak-OTFS) and avoiding interference (CP-OFDM). Standards discussions typically focus on low mobility / small cells characteristic of North America, but there are important wireless markets where large cells are the rule rather than the exception. Our analysis provides insight into how well technology developed for one market transfers to market with different characteristics. It also provides insight into SE as we start to make use of higher frequency bands where higher Doppler shifts are more common.

Section \ref{seccpofdm} introduces CP-OFDM, where the carrier waveform is a sinusoid with frequency an integer multiple of the sub-carrier spacing (SCS) $\Delta f$. The information grid is defined in the frequency domain (FD), the number of carriers is $B/\Delta f$, where $B$ is the bandwidth, and the time duration $T = 1/\Delta f$. The carriers are pairwise orthogonal, enabling low-complexity per sub-carrier equalization at the receiver \cite{OFDM1971}.
CP-OFDM performance is however sensitive to high channel delay and Doppler spread \cite{Wang2006}.

Section \ref{seczakotfs} introduces Zak-OTFS, where the carrier waveform is a pulse in the delay-Doppler (DD) domain, that is a quasi-periodic localized function defined by a delay period $\tau_p$ and a Doppler period $\nu_p = 1/\tau_p$. When viewed in the time-domain (TD) this function is realized as a pulse train modulated by a tone, hence the name pulsone. The time duration ($T$) and bandwidth ($B$) of a pulsone are inversely proportional to the characteristic width of the DD domain pulse along the Doppler axis and the delay axis respectively. The number of non-overlapping DD domain pulses, each spread over an area $1/BT$, is equal to the time-bandwidth product $BT$. For more details see \cite{zakotfs1, zakotfs2}.

Both CP-OFDM and Zak-OTFS can allocate orthogonal TF resources to different users. In Zak-OTFS this is accomplished by choosing an appropriate pulse shaping filter, and we refer the reader to \cite{MUZakOTFS} for details. In Section \ref{sec5} we describe how CP-OFDM introduces a cyclic prefix (CP) to prevent successive OFDM symbols from interfering in time. The CP duration is required to be at least the channel delay spread $\tau_s$. In Zak-OTFS, we may introduce a guard time of duration $\tau_s$ between consecutive subframes. CP-OFDM has higher overhead because there are multiple OFDM symbols per subframe. We also discuss the characteristics of FD interference from adjacent users.

The choice between CP-OFDM and Zak-OTFS is the choice between avoiding interference between carriers/symbols (ICI/ISI) and embracing/equalizing interference (see Section \ref{sec5}). In the presence of channel Doppler spread, CP-OFDM cannot efficiently acquire the complete I/O relation (i.e., coefficients of the interference between sub-carriers) and it is therefore extremely challenging for it to equalize the inter-carrier interference (ICI). CP-OFDM therefore chooses its SCS to be significantly greater than the channel Doppler spread and the symbol duration to be significantly greater than the CP duration. By choosing such an information grid much coarser than the physical channel we are able to limit CP overhead and avoid ICI. 

In Section \ref{sec5} we describe how Zak-OTFS I/O relation can be acquired for every carrier waveform (i.e., complete acquisition) from the response to a single pilot carrier waveform (i.e., efficient acquisition), as long as the Zak-OTFS delay and Doppler period parameters are chosen to be greater than the channel delay and Doppler spread respectively (a condition referred to as the crystallization condition). 
For doubly-spread channels, the carrier spacing $1/B$ in delay is less than the channel delay spread, and the carrier spacing $1/T$ is less than the channel Doppler spread. The choice of such a fine information grid embraces ICI, and the Zak-OTFS carriers are designed to make joint equalization possible by interfering with each other in the same way.

Another important aspect of comparison between Zak-OTFS and CP-OFDM discussed in Section \ref{sec5} is that, in CP-OFDM the accuracy and efficiency of acquisition of the I/O relation is coupled with the amount of channel delay and Doppler spread. A higher Doppler spread necessitates a higher pilot density along time due to the I/O relation being time-selective. For Zak-OTFS, accuracy and efficiency of acquisition of the I/O relation is however decoupled with the amount of channel delay and Doppler spread, in the sense that it is possible to acquire the I/O relation with low pilot overhead (efficient acquisition) as long as the product of the maximum possible channel delay and Doppler spread is significantly less than $1/4$ (valid for most 6G scenarios).   

Section \ref{secopoint} discusses the choice of operating point. The SCS $\Delta f$ determines the operating point for CP-OFDM, with $\Delta f$ increasing when the Doppler spread is high, and decreasing when the delay spread is high. Once the SCS is fixed, there are still many possible variations within the 3GPP standard, and the optimization of SE in Section \ref{sec7} takes all these variations into account. With Zak-OTFS there are more free parameters, and Section \ref{secopoint} describes how to choose (i) the delay and Doppler periods; (ii) the transmit pulse shaping filter; and (iii) the pilot and guard regions. These choices are informed by numerical simulations for the Veh-A channel model \cite{EVAITU} which comprises six channel paths and is representative of real propagation environments. When we optimize SE, we can trade the overhead of the pilot/guard region for less Doppler aliasing in the pilot region and more accurate acquisition of the I/O relation. Section \ref{secopoint} discusses how to optimize Zak-OTFS SE by balancing these two effects.

SE depends on mobility (high/low) and on cell size (large/small) and in Section \ref{sec7}, we overlay $35$ combinations of Doppler and delay spread on these four quadrants. We optimize SE for CP-OFDM over all variants specified in the 3GPP 5G NR standard, and we optimize SE for Zak-OTFS over the operating points discussed in Section \ref{secopoint}. Each of the above $35$ combinations identifies a cell in the delay-Doppler grid and is labeled by the relative SE of Zak-OTFS and CP-OFDM. While Zak-OTFS and CP-OFDM achieve similar spectral efficiencies for low mobility small cell scenarios, we show that Zak-OTFS shows significant gains when either the delay spread, or the Doppler spread is high.

\section{System model}
We consider a doubly-spread channel characterized by a delay-Doppler (DD) spreading function $h_{\text{phy}}(\tau, \nu)$ such that the received TD signal is given by\cite{Bello}

{\vspace{-4mm}
\small
\begin{eqnarray}
\label{eqn1}
    y(t) & \hspace{-3mm} = & \hspace{-3mm} \iint h_{\text{phy}}(\tau, \nu) \, x(t - \tau) \, e^{j 2 \pi \nu (t - \tau)} \, d\tau \, d\nu \, + n(t),
\end{eqnarray}\normalsize}where $x(t)$ is the transmitted TD signal and $n(t)$ is AWGN at the receiver.
The transmitted signal $x(t)$ is approximately time and bandwidth limited to $T$ seconds and $B$ Hz respectively. As an example, if the channel consists of $L$ paths, with the $i$-th path having channel gain $h_i$, delay $\tau_i$ and Doppler shift $\nu_i$ then

{\small
\vspace{-4mm}
\begin{eqnarray}
\label{eqn2}
    h_{\text{phy}}(\tau, \nu) & = & \sum\limits_{i=1}^L h_i \, \delta(\tau - \tau_i) \, \delta(\nu - \nu_i)
\end{eqnarray}\normalsize}and using (\ref{eqn1}) the received signal is given by

{\vspace{-4mm}
\small
\begin{eqnarray}
\label{eqn3}
    y(t) & = & \sum\limits_{i=1}^L h_i \, x(t - \tau_i) \, e^{j 2 \pi \nu_i (t - \tau_i)} \, + \, n(t).
\end{eqnarray}\normalsize}

\section{CP-OFDM}
\label{seccpofdm}
For CP-OFDM, let $X[k], k=0,1,\cdots BT -1$ denote the $BT$ information symbols in a symbol.
The transmitted CP-OFDM signal is given by \cite{OFDM1971}

{\vspace{-4mm}
\small
\begin{eqnarray}
\label{eqn4}
    x(t) & = \begin{cases}
        \frac{1}{\sqrt{T}} \, \sum\limits_{k=0}^{BT-1} X[k] \, e^{j 2 \pi k \Delta f t} \, &, \, -T_{\text{cp}} \, < t < T \\
        0 \, &, \, \text{otherwise} \\
     \end{cases},
\end{eqnarray}\normalsize}where $\Delta f \Define 1/T$ is the sub-carrier spacing (SCS) and $T_{cp}$ is the CP duration. Note that the symbol duration is $(T + T_{\text{cp}})$ and the $k$-th information symbol $X[k]$ is carried by the $k$-th carrier waveform (called as the $k$-th sub-carrier) which is a sinusoid of frequency $k \Delta f$. The frequency spacing between adjacent carriers is $\Delta f$. Also, since $e^{j 2 \pi k \Delta f t}$ is periodic with period $T$, the part of $x(t)$ for $-T_{\text{cp}} < t < 0$ is exactly same as $x(t)$ for $T - T_{\text{cp}} < t < T$, and is appropriately called as the CP. The $BT$ carriers are orthogonal, i.e.,
the inner product between any two distinct carrier waveform is zero, i.e., $\int_{0}^T e^{j 2 \pi k \Delta f t} \, e^{-j 2 \pi l \Delta f t} \, dt = 0$, for all $k \ne l$. 

Substituting (\ref{eqn4}) into (\ref{eqn3}) gives the expression of the received signal $y(t)$ in terms of the information symbols $X[k], k=0,1,\cdots, BT -1$. The signal received on the $m$-th sub-carrier, $m=0,1,\cdots, BT -1$ is given by

{\vspace{-4mm}
\small
\begin{eqnarray}
\label{eqn5}
    Y[m] & = & \frac{1}{\sqrt{T}} \int\limits_{0}^T y(t) \, e^{-j 2 \pi m \Delta f t} \, dt. 
\end{eqnarray}\normalsize}Substituting $y(t)$ in terms of $X[k]$ into the R.H.S. above gives the frequency domain (FD) I/O relation\cite{SKOTFS2}

{\vspace{-4mm}
\small
\begin{eqnarray}
\label{eqn6}
    Y[m] & = & X[m] \, H_{\text{ofdm}}[m,m] \nonumber \\
    & & + \underbrace{\sum\limits_{k=0, k \ne m}^{BT -1} X[k] \, H_{\text{ofdm}}[m,k]}_{\text{Inter-carrier interference (ICI)}} \, + \, N[m], \nonumber \\
    H_{\text{ofdm}}[m,k] & \Define & \sum\limits_{i=1}^L {\Big [} h_i \, e^{-j 2\pi \frac{\tau_i}{T}(\nu_i T + k)} \, e^{j \pi (\nu_i T + k - m)} \nonumber \\
    & & \hspace{9mm} \text{sinc}(\nu_i T + k -m ) {\Big ]},
\end{eqnarray}\normalsize}where $N[m] = \frac{1}{\sqrt{T}} \int\limits_{0}^T n(t) \, e^{-j 2 \pi m \Delta f t} \, dt, m=0,1,\cdots, BT-1$ are i.i.d. zero mean complex Gaussian AWGN samples.
In the special case where there is no Doppler shift $\nu_i =0, i=1,2,\cdots, L$, then $H_{\text{ofdm}}[m,k] = 0$ for all $m \ne k$, i.e., there is no inter-carrier interference (ICI) term in (\ref{eqn6}), $Y[m] = H_{\text{ofdm}}[m,m] x[m] + N[m]$ and hence per-carrier equalization suffices. However, in the presence of Doppler shifts, the amount of ICI depends on $\nu_i$. From the term $\text{sinc}(\nu_i T + k - m)$ in (\ref{eqn6}) it is clear that ICI is small if $\nu_i T \ll 1$, i.e., $\nu_i \ll \Delta f$. Therefore, per-carrier equalization suffices if

{\vspace{-4mm}
\small
\begin{eqnarray}
\label{eqn7}
\Delta f & \gg & \max_i \nu_i. 
\end{eqnarray}\normalsize}

\subsection{Time and frequency selectivity}
For a doubly-spread channel with DD spreading function $h_{\text{phy}}(\tau, \nu)$,
the received signal $y(t)$ is also equivalently expressed as\cite{LTVbook}

{\vspace{-4mm}
\small
\begin{eqnarray}
\label{eqn8}
        y(t)
        \hspace{-3mm} & = & \hspace{-3mm} \int  H(t,f) \, X(f) \, e^{j 2 \pi ft} \, df \, + n(t), \nonumber \\
        H(t,f) & \Define & \iint h_{\text{phy}}(\tau, \nu) \, e^{-j2 \pi \nu \tau} \, e^{j 2 \pi (\nu t - f \tau)} \, d\tau \, d\nu, \nonumber \\
        & = & \sum\limits_{i=1}^L h_i \, e^{-j 2 \pi \nu_i \tau_i} \, \underbrace{e^{j 2 \pi \nu_i t}}_{\text{Time variation}} \, \underbrace{e^{-j 2 \pi \tau_i f}}_{\text{Freq. variation}}
\end{eqnarray}\normalsize}where $X(f)$ is the Fourier transform of $x(t)$. $H(t,f)$ is the time-frequency representation of the channel. For example, the TD channel response to the $k$-th carrier waveform $x(t) = e^{j 2 \pi k \Delta f t}$ will be $H(t,k\Delta f) \, e^{j 2 \pi k \Delta f t}$. In other words, the channel response varies in both time and frequency, i.e., it is both time-selective and frequency-selective. Hence the I/O relation in (\ref{eqn6}) varies from one CP-OFDM symbol to the next (i.e., $H_{\text{ofdm}}[m,k]$ varies from one symbol to the next). Due to the term $e^{j 2 \pi \nu_i t}$, the channel response along the $i$-th path rotates by $2 \pi$ in every $1/\nu_i$ seconds. Similarly, due to the term $e^{-j 2 \pi f \tau_i}$, the channel response along the $i$-th path rotates by $2 \pi$ in every $1/\tau_i$ Hz. Hence the magnitude of the channel response in time (proportional to received signal power) is approximately stationary only over a coherence time-interval $T_c \ll \frac{1}{\max_i \nu_i - \min_i \nu_i}$. Similarly, the magnitude of the channel response in frequency domain (proportional to received signal power) is approximately stationary only over a coherence bandwidth $B_c \ll \frac{1}{\max_i \tau_i - \min_i \tau_i}$, i.e.

{\vspace{-4mm}
\small
\begin{eqnarray}
\label{eqn9}
    T_c & \ll & \frac{1}{\nu_s} \,,\, B_c \ll \frac{1}{\tau_s},
\end{eqnarray}\normalsize}where $\tau_s$ and $\nu_s$ denote the  channel delay and Doppler spread, i.e.

{\vspace{-4mm}
\small
\begin{eqnarray}
\label{eqn10}
    \nu_s & \Define & \max_i \nu_i - \min_i \nu_i, 
    \tau_s  \Define  \max_i \tau_i - \min_i \tau_i.
\end{eqnarray}\normalsize}
\section{Zak-OTFS}
\label{seczakotfs}
In the following we summarize Zak-OTFS modulation and demodulation and its I/O relation (see \cite{zakotfs1, zakotfs2, otfsbook} for a detailed exposition).
In Zak-OTFS modulation, information is carried by quasi-periodic pulses in the DD domain which are
periodic with period $\nu_p$ along the Doppler axis and quasi-periodic with period $\tau_p = 1/\nu_p$ along the delay axis. Let $M \Define B \tau_p, N \Define T \nu_p$ be integers.
Also, let $x[k,l], k=0,1,\cdots, M-1$, $l=0,1,\cdots, N-1$ denote the $MN = BT$ information symbols. Then the discrete DD domain quasi-periodic information signal is given by

{\vspace{-4mm}
\small
\begin{eqnarray}
\label{eqn11}
    x_{dd}[k,l] & = & e^{j 2 \pi \lfloor \frac{k}{M} \rfloor \frac{l}{N}} \, x[k \, \text{mod} \, M \, , \, l \, \text{mod} \, N],
\end{eqnarray}\normalsize}$k, l \in {\mathbb Z}$. This implies that for any $n,m, k, l \in {\mathbb Z}$

{\vspace{-4mm}
\small
\begin{eqnarray}
\label{eqn12}
    x_{dd}[k + nM, l + mN] & = & e^{j 2 \pi n \frac{l}{N}} \, x_{dd}[k,l],
\end{eqnarray}\normalsize}i.e., $x_{dd}[k,l]$ is periodic along the discrete Doppler axis with period $N$ and
is quasi-periodic along the discrete delay axis with period $M$. The discrete DD symbols $x_{dd}[k,l], k,l \in {\mathbb Z}$ are then carried by Dirac-delta pulses located at points of the information lattice $\Lambda = \{ (\tau, \nu) = (k/B, l/T) \, | \, k,l \in {\mathbb Z}\}$ in the continuous DD domain. In other words, the corresponding continuous DD domain signal is given by

{\vspace{-4mm}
\small
\begin{eqnarray}
\label{eqn13}
    x_{dd}(\tau, \nu) & \hspace{-3mm} = & \hspace{-3mm} \sum\limits_{k,l \in {\mathbb Z}} x_{dd}[k,l] \delta\left(\tau - \frac{k}{B}\right) \, \delta\left(\nu - \frac{l}{T} \right).
\end{eqnarray}\normalsize}Note that $x_{dd}(\tau, \nu)$ is quasi-periodic with respect to the period lattice
$\Lambda_p = \{ (\tau, \nu) = (n \tau_p , m \nu_p) \, | \, n,m \in {\mathbb Z}\}$, i.e., for $n,m \in {\mathbb Z}$

{\vspace{-4mm}
\small
\begin{eqnarray}
\label{eqn14}
    x_{dd}(\tau + n \tau_p, \nu + m \nu_p) & = & e^{j 2 \pi n \tau \nu_p} \, x_{dd}(\tau, \nu).
\end{eqnarray}\normalsize}The TD realization of a DD domain signal is given by its inverse Zak transform.
For $x_{dd}(\tau, \nu)$ in (\ref{eqn13}), its TD realization has infinite time and bandwidth and therefore it is filtered (pulse-shaping) with a transmit pulse $w_{tx}(\tau, \nu)$ resulting in
the pulse shaped signal

{\vspace{-4mm}
\small
\begin{eqnarray}
\label{eqn15}
    x_{dd}^{w_{tx}}(\tau, \nu) & = & w_{tx}(\tau, \nu) \, *_{\sigma} \, x_{dd}(\tau, \nu),
\end{eqnarray}\normalsize}where the spread of the pulse $w_{tx}(\tau, \nu)$ is approximately $1/B$ and $1/T$ along the delay and Doppler axis respectively (i.e., inversely proportional to bandwidth $B$ and time $T$). Here $*_{\sigma}$ denotes the twisted convolution operation and is given by

{\vspace{-4mm}
\small
\begin{eqnarray}
\label{eqn16}
    w_{tx}(\tau, \nu) \, *_{\sigma} \, x_{dd}(\tau, \nu) & & \nonumber \\
 & & \hspace{-40mm} = \iint w_{tx}(\tau', \nu') \, x_{dd}(\tau - \tau', \nu - \nu') \, e^{j 2 \pi \nu' (\tau - \tau')} \, d\tau' \, d\nu'. \nonumber \\
\end{eqnarray}\normalsize}Since twisted convolution preserves quasi-periodicity, $x_{dd}^{w_{tx}}(\tau, \nu)$
is quasi-periodic w.r.t. $\Lambda_p$. The TD transmit signal is given by the TD representation of $x_{dd}^{w_{tx}}(\tau, \nu)$, i.e., its Inverse Zak-transform

{\small
\vspace{-4mm}
\begin{eqnarray}
\label{eqn17}
    x(t)  =  {\mathcal Z}_t^{-1}\left( x_{dd}^{w_{tx}}(\tau, \nu) \right) 
     =  \sqrt{\tau_p} \int\limits_{0}^{\nu_p} x_{dd}(t, \nu) \, d\nu.
\end{eqnarray}\normalsize}At the receiver, the continuous DD domain representation of $y(t)$ is given by its Zak transform

{\vspace{-4mm}
\small
\begin{eqnarray}
\label{eqn18}
    y_{dd}(\tau, \nu) & = & {\mathcal Z}_t(y(t)) \nonumber \\
    & =  & \sqrt{\tau_p} \sum\limits_{n \in {\mathbb Z}} y(\tau + n \tau_p) \, e^{-j 2 \pi n \nu \tau_p}.
\end{eqnarray}\normalsize}Channel action in DD domain is described through twisted convolution, i.e. from (\ref{eqn1}), (\ref{eqn17}) and (\ref{eqn18}) it follows that

{\vspace{-4mm}
\small
\begin{eqnarray}
\label{eqn19}
    y_{dd}(\tau, \nu) & \hspace{-3mm} = & \hspace{-3mm} h_{\text{phy}}(\tau, \nu) \, *_{\sigma} \, x_{dd}^{w_{tx}}(\tau, \nu) \, + \, n_{dd}(\tau, \nu),
\end{eqnarray}\normalsize}where $n_{dd}(\tau, \nu) = {\mathcal Z}_t(n(t))$ is the DD domain representation of the AWGN $n(t)$.
$y_{dd}(\tau, \nu)$ is then match-filtered with a receive pulse $w_{rx}(\tau, \nu)$ resulting
in the match-filtered received signal

{\vspace{-4mm}
\small
\begin{eqnarray}
\label{eqn20}
    y_{dd}^{w_{rx}}(\tau, \nu) & = & w_{rx}(\tau, \nu) \, *_{\sigma} \, y_{dd}(\tau, \nu).
\end{eqnarray}\normalsize}We consider $w_{rx}(\tau, \nu) = w_{tx}^*(-\tau, -\nu) \, e^{j 2 \pi \nu \tau}$ as it optimizes the received signal-to-noise ratio \cite{Hanly2024}. From (\ref{eqn15}), (\ref{eqn19}) and (\ref{eqn20}) it follows that the AWGN-free part of $y_{dd}^{w_{rx}}(\tau, \nu)$ is related to the information signal $x_{dd}(\tau, \nu)$ through the simple relation

{\vspace{-4mm}
\small
\begin{eqnarray}
\label{eqn21}
y_{dd}^{w_{rx}}(\tau, \nu) & = & h_{\text{eff}}(\tau, \nu) \, *_{\sigma} \,  x_{dd}(\tau, \nu),
\end{eqnarray}\normalsize}where the DD domain effective channel filter $h_{\text{eff}}(\tau, \nu)$
is given by

{\vspace{-4mm}
\small
\begin{eqnarray}
    \label{eqn22}
    h_{\text{eff}}(\tau, \nu) & \hspace{-3mm} = &  \hspace{-3mm} w_{rx}(\tau, \nu) \, *_{\sigma} \, h_{\text{phy}}(\tau, \nu) \, *_{\sigma} \, w_{tx}(\tau, \nu).
\end{eqnarray}\normalsize}This follows due to the fact that the twisted convolution operation is associative.
Finally, sampling $y_{dd}^{w_{rx}}(\tau, \nu)$ on the information lattice $\Lambda$ gives the received discrete DD domain signal

{\vspace{-4mm}
\small
\begin{eqnarray}
    \label{eqn23}
    y_{dd}[k,l] & \Define & y_{dd}^{w_{rx}}\left(\tau = \frac{k}{B} \,,\, \nu = \frac{l}{T} \right) \nonumber \\
    & & \hspace{-7mm} = h_{\text{eff}}[k,l] \, *_{\sigma} \, x_{dd}[k,l] \, + \, n_{dd}[k,l], \, \text{where}
\end{eqnarray}\normalsize}

{\vspace{-6mm}
\small
\begin{eqnarray}
    \label{eqn24}
    h_{\text{eff}}[k,l] & \Define & h_{\text{eff}}\left(\tau = \frac{k}{B} \,,\, \nu = \frac{l}{T} \right),
\end{eqnarray}\normalsize}$k,l \in {\mathbb Z}$. $h_{\text{eff}}[k,l]$ is the discrete DD domain effective channel filter and $*_{\sigma}$ here denotes discrete twisted convolution

{\vspace{-4mm}
\small
\begin{eqnarray}
\label{eqn25}
    h_{\text{eff}}[k,l] \, *_{\sigma} \, x_{dd}[k,l] & & \nonumber \\
    & & \hspace{-36mm} = \sum\limits_{k', l' \in {\mathbb Z}} h_{\text{eff}}[k',l'] \, x_{dd}[k - k' , l - l'] \, e^{j 2 \pi l' \frac{(k - k')}{MN}}.
\end{eqnarray}\normalsize}The delay and Doppler axis spread of $h_{\text{phy}}(\tau, \nu)$ is $\tau_s$ and $\nu_s$ respectively, and therefore from (\ref{eqn22}) it follows that the delay and Doppler axis spread of $h_{\text{eff}}(\tau, \nu)$ is approximately $(\tau_s + O(1/B))$ and $(\nu_s + O(1/T))$ respectively. Hence, from (\ref{eqn24}), the spread of $h_{\text{eff}}[k,l]$ along the discrete
delay and Doppler axis is approximately $B \tau_s$ and $T \nu_s$ respectively.
\begin{figure}[!h]
\centering
\includegraphics[width=8.4cm, height=6.0cm]{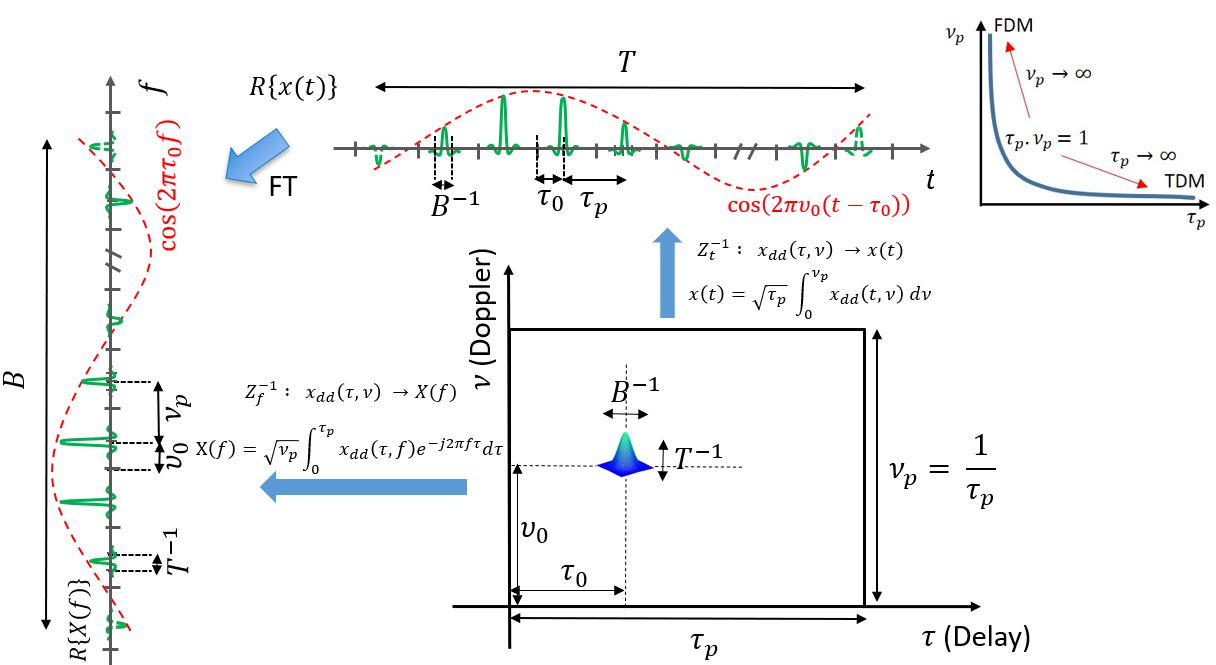}
\caption{{A quasi-periodic DD domain pulse and its TD (FD) realization, referred to as TD (FD) pulsone.}} 
\label{fig_4}
\end{figure}
In CP-OFDM, an information symbol is carried in a symbol duration over a frequency sub-carrier,
and its energy is therefore localized in a time-frequency interval $T \times \Delta f$. On the other hand, in Zak-OTFS each information symbol is spread through the entire bandwidth $B$ and also time duration $T$. Towards this end, note that using (\ref{eqn11}), (\ref{eqn13}) can also be expressed in terms of the information symbols $x[k,l]$ as

{\vspace{-4mm}
\small
\begin{eqnarray}
\label{eqn26}
    x_{dd}(\tau, \nu) & = & \sum\limits_{k=0}^{M-1} \sum\limits_{l=0}^{N-1} x[k,l] x_{dd}(\tau, \nu ; k/B, l/T),
\end{eqnarray}\normalsize}where $x_{dd}(\tau, \nu ; k/B, l/T)$ is the quasi-periodic DD domain Dirac-delta pulse at $(\tau_0, \nu_0) = (k/B, l/T)$ which carries $x[k,l]$. In general, the quasi-periodic DD domain Dirac-delta pulse at $(\tau_0, \nu_0)$ is given by \cite{otfsbook}

{\vspace{-4mm}
\small
\begin{eqnarray}
\label{eqn27}
    x_{dd}(\tau, \nu ; \tau_0, \nu_0) & & \nonumber \\
    & & \hspace{-33mm} = \sum\limits_{n \in {\mathbb Z}} \sum\limits_{m \in {\mathbb Z}} e^{j 2 \pi n \nu_0 \tau_p} \, \delta(\tau - \tau_0 - n \tau_p) \, \delta(\nu - \nu_0 - m \nu_p).
\end{eqnarray}\normalsize}Let

{\small
\vspace{-4mm}
\begin{eqnarray}
\label{eqn281}
    p_{\tau_0, \nu_0}^w(t) & \Define & {\mathcal Z}_t^{-1}\left( w_{tx}(\tau, \nu) *_{\sigma} x_{dd}(\tau, \nu ; \tau_0, \nu_0) \right).
\end{eqnarray}\normalsize}Then, from (\ref{eqn15}), (\ref{eqn17}), (\ref{eqn26}) and (\ref{eqn281}) it follows that

{\small
\vspace{-4mm}
\begin{eqnarray}
    x(t) & = & \sum\limits_{k=0}^{M-1} \sum\limits_{l=0}^{N-1} x[k,l] \, p_{\frac{k}{B}, \frac{l}{T}}^w(t), 
\end{eqnarray}\normalsize}i.e., $x[k,l]$ is carried by the TD carrier waveform $p_{k/B, l/T}^w(t)$ which is the TD realization of a pulse shaped quasi-periodic DD domain Dirac-delta function located at $(k/B, l/T)$ on the information lattice $\Lambda$.
For a factorizable transmit pulse $w_{tx}(\tau, \nu) = W_1(\tau) W_2(\nu)$, the Zak-OTFS carrier waveform for a DD Dirac-delta pulse at $(\tau_0, \nu_0)$ is given by \cite{otfsbook}

{\small
\vspace{-4mm}
\begin{eqnarray}
\label{eqn28}
    p_{\tau_0, \nu_0}^w(t) & = &  \sqrt{\tau_p} \sum\limits_{n= - \infty}^{\infty}  g(n \tau_p + \tau_0) \,  W_1(t - n \tau_p - \tau_0), \nonumber \\
    & & \hspace{-25mm} g(t) \Define w_2(t) \, e^{j 2 \pi \nu_0 (t - \tau_0)} \,,\,  w_2(t) \Define  \int W_2(\nu) \, e^{j 2 \pi \nu t} \, d\nu.
\end{eqnarray}\normalsize}The frequency domain realization of this carrier waveform corresponding to a quasi-periodic Dirac-delta DD pulse at $(\tau_0, \nu_0)$ is given by

{\vspace{-4mm}
\small
\begin{eqnarray}
\label{eqn29}
    P_{\tau_0, \nu_0}^w(f) & = & \int p_{\tau_0, \nu_0}^w(t) \, e^{-j 2 \pi f t} \, dt\nonumber \\
    & & \hspace{-21mm} = \left[ \sqrt{\nu_p} \sum\limits_{m= - \infty}^{\infty}  \hspace{-3mm} e^{-j 2 \pi \tau_0 (m \nu_p + \nu_0)} \,  W_2(f - m \nu_p - \nu_0)  \right] w_1(f) \, \nonumber \\
     & & \hspace{-20mm} w_1(f) \Define \int W_1(t) \, e^{-j 2 \pi f t} \, dt.
\end{eqnarray}\normalsize}In (\ref{eqn28}), $w_2(t)$ limits $p_{\tau_0, \nu_0}^w(t)$ in time, and in (\ref{eqn29}), $w_1(f)$ limits $P_{\tau_0, \nu_0}^w(f)$ in bandwidth. As an example with sinc pulse shaping

{\vspace{-4mm}
\small
\begin{eqnarray}
\label{eqn30}
    W_1(\tau) =  \sqrt{B} \text{sinc}(B \tau) \,,\, W_2(\nu) = \sqrt{T} \text{sinc}(T \nu),
\end{eqnarray}\normalsize}$w_2(t)$ and $w_1(f)$ are rectangular waveforms limited to $T$ seconds and $B$ Hz respectively. In (\ref{eqn28}), the TD carrier waveform $p_{\tau_0, \nu_0}^w(t)$ is a pulse train with TD pulses at $t = (n \tau_p + \tau_0), n \in {\mathbb Z}$, modulated by $g(t)$ which is a tone of frequency $\nu_0$ time-limited to $T$ seconds. The carriers waveforms are therefore called \emph{pulsones} (pulse train modulated by a tone). Interestingly, in (\ref{eqn29}), the FD representation of a TD pulsone is also a FD pulse train with pulses at $f = m\nu_p + \nu_0 $, $m \in {\mathbb Z}$, modulated by a FD tone $e^{-j 2 \pi \tau_0 f}$ (a FD pulsone). The energy of a pulsone carrier is therefore spread throughout the allocated TF resource. An illustration of the TD, FD and DD domain representations of the Zak-OTFS carrier waveform is illustrated through Fig.~\ref{fig_4}.


Also, each choice of period parameters $(\tau_p, \nu_p)$ on the rectangular hyperbola $\tau_p \, \nu_p = 1$ gives a new Zak-OTFS modulation. When $\tau_p \rightarrow \infty$ and $\nu_p \rightarrow 0$, the TD pulsone has only one TD pulse at $t=\tau_0$ and Zak-OTFS becomes same as TD modulation (TDM). Similarly, when $\nu_p \rightarrow \infty$ and $\tau_p \rightarrow 0$, the FD realization of the pulsone is a single pulse at $f = \nu_0$ and Zak-OTFS becomes FD modulation (FDM). Therefore, Zak-OTFS is a generalization of TDM and FDM.

Note that practical implementation of the Zak-OTFS transceiver described above is achieved through Discrete Zak transforms as described in \cite{otfsbook, zakofdm}.


   \begin{figure}
     \centering
        \vspace{-9mm}
        \includegraphics[width=7.52cm,height=5.0cm]{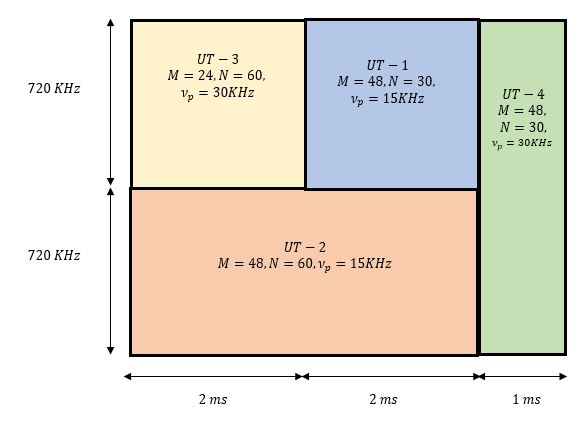}
        \vspace{-2mm}
        \caption{Allocation of TF resources.}
        \label{fig1}
        \vspace{-3mm}
    \end{figure}
    
\section{System overheads}
\label{sec5}
The ability to allocate orthogonal TF resources to users is a strength
of CP-OFDM. Zak-OTFS is able to match this strength by choosing an
appropriate pulse shaping filter $w_{tx}(\tau, \nu)$. For details, see \cite{MUZakOTFS}, and for an illustrative example see Fig.~\ref{fig1}. 

\subsection{Managing interference between users}
CP-OFDM introduces a cyclic prefix (CP) to prevent successive OFDM
symbols from interfering in time. The CP duration $T_{cp}$ is required
to be at least the channel delay spread $\tau_s$ for a symbol period
$T_{prb} = T_{cp} + 1/\Delta f$, where $\Delta f$ is the SCS. In Zak-OTFS we may introduce a guard time of duration
$\tau_s$ between consecutive subframes. CP-OFDM has higher overhead because there are multiple OFDM symbols per subframe.

\uline{Example:} CP-OFDM and zak-OTFS with $1$ ms subframes, channel
delay spread $\tau_s = 4.7 \, \mu s$. If the SCS $\Delta f = 15$ kHz
in CP-OFDM, then the CP overhead $T_{cp}/T_{prb}$ is about $6.6 \%$.
In Zak-OTFS, the guard time overhead $\tau_s/ (T + \tau_s)$ is only about $0.47 \, \%$.

Adjacent users also introduce FD interference. In Zak-OTFS this interference is equally distributed over all $MN$ pulsone carriers (since the energy of a pulsone carrier is spread throughout the TF domain, see paragraph after (\ref{eqn30})).  On the other hand, in CP-OFDM adjacent user interference is concentrated at the boundary sub-carriers.
If guard bands in time and frequency were omitted, we would not expect
significant degradation
in Zak-OTFS performance.

\subsection{Acquiring the input-output (I/O) relation}
\label{subseciorel2}
When we consider CP-OFDM in the absence of Doppler shifts, it follows
from (\ref{eqn6}) that there is no ICI, and the off-diagonal coefficients $H_{\text{ofdm}}[m,k] = 0$ for all $m \ne k$. We can dedicate a single CP-OFDM symbol for channel acquisition and use that
symbol to transmit pilots  on all sub-carriers. Since there is no Doppler shift, the diagonal coefficients $H_{\text{ofdm}}[m,m]$, $m=0,1,\cdots, BT-1$, do not change from one symbol to the next, and we can use the pilot responses to equalize the sub-carriers. The overhead of a single pilot symbol can be amortized over the sub-frame
and is not significant.


 However, when we consider CP-OFDM in the presence of Doppler shifts, there is time-selectivity, and the coefficients $H_{\text{ofdm}}[m,k]$ are time-varying. The off-diagonal coefficients $H_{\text{ofdm}}[m,k]$ need not be zero, and the problem of estimating
 $B^2T^2$ coefficients from pilots on BT sub-carriers becomes intractable. In practice, we transmit pilots only on predetermined 
 symbols and predetermined sub-carriers and we interpolate the responses to estimate $H_{\text{ofdm}}[m,m]$ for all sub-carriers
 in all symbols of the subframe. This however ignores estimation of the ICI coefficients $H_{\text{ofdm}}[m,k]$, $m \ne k$ and therefore acquisition of the I/O relation is incomplete. These pilots are referred to as DeModulation Reference Signals (DMRS) in 3GPP.
 When ICI and Gaussian noise are commensurate, we can perform per-carrier equalization accurately using estimates for the diagonal coefficients $H_{\text{ofdm}}[m,m]$.  When ICI is more significant, it degrades performance.


We now turn to Zak-OTFS. We recall from \cite{zakotfs2} that when the delay period of the Zak-OTFS carrier is greater than the delay spread of the channel, and the Doppler period of the carrier is greater than the Doppler spread of the channel, the I/O relation can simply be read off from the response to a single pilot carrier waveform at DD location $(k_p, l_p)$. The condition 
\begin{eqnarray}
\label{eqn31}
    \tau_p & > & \tau_s \,,\, \nu_p \, > \, \nu_s
\end{eqnarray}is called the crystallization condition. We acquire the I/O relation by acquiring coefficients $h_{\text{eff}}[k,l]$ for all $(k,l)$ in the support set ${\mathcal H}$. The crystallization condition implies that translates of ${\mathcal H}$ by points $(nM, mN)$ in the period lattice do not overlap. The discrete DD domain pilot signal satisfying the quasi-periodicity property in (\ref{eqn12}) is given by 

{\small
\vspace{-4mm}
\begin{eqnarray}
\label{eqn32}
    x_{p,dd}[k,l]  = \hspace{-3mm} \sum\limits_{n,m \in {\mathbb Z}} \hspace{-3mm} e^{j 2 \pi \frac{n l}{N}} \delta[k - k_p - nM] \delta[l - l_p - mN],
\end{eqnarray}\normalsize}and in the absence of noise, the response is given by (see (\ref{eqn23}))

{\vspace{-4mm}
\small
\begin{eqnarray}
\label{eqn33}
    y_{p,dd}[k,l] & = & h_{\text{eff}}[k,l] *_{\sigma} x_{p,dd}[k,l] \nonumber \\
    &  & \hspace{-20mm} = {h_{\text{eff}}[k - k_p,l - l_p] \, e^{j 2 \pi k_p \frac{(l - l_p)}{MN}}} \nonumber \\
& & \hspace{-19mm} + \hspace{-3mm} \sum\limits_{\substack{n,m \in {\mathbb Z} \\ (n,m) \ne (0,0)}}
\hspace{-4mm} {\Big (} h_{\text{eff}}[k - k_p - nM,l - l_p -mN] e^{j 2 \pi n \frac{l}{N}} \nonumber \\
& & e^{j 2 \pi (l  - l_p - mN) \frac{(k_p + nM)}{MN}} {\Big )}.
\end{eqnarray}\normalsize}
   \begin{figure}
     \centering
        \vspace{-5mm}
        \includegraphics[width=7.72cm,height=5.0cm]{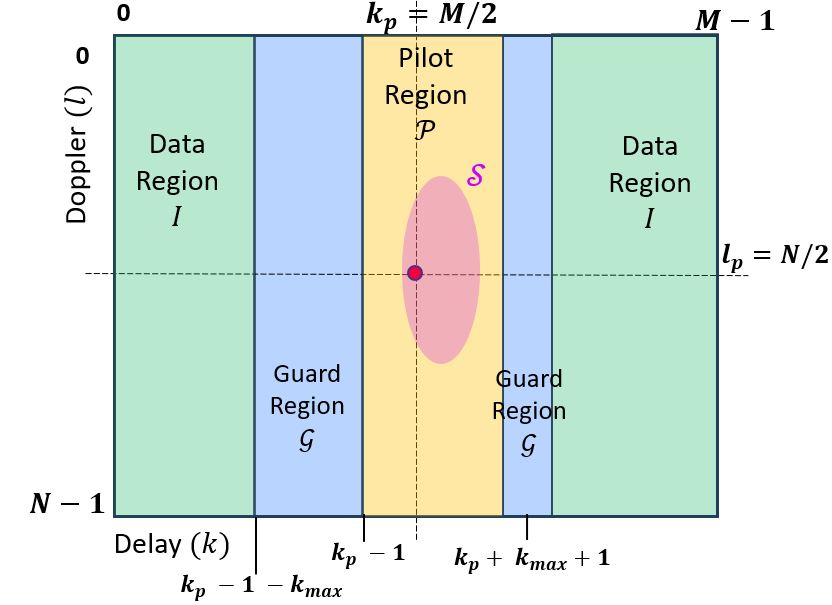}
        \vspace{-2mm}
        \caption{An illustrative Zak-OTFS sub-frame with pilot, guard and data regions. $k_{max} \Define \lceil B \tau_s \rceil$ and the fractional overhead of the pilot and guard regions is $(2 k_{max} + 3)/M \approx 2 \tau_s /\tau_p$. Note that with $\tau_p \gg \tau_s$, the pilot overhead is small, and the I/O relation can be acquired efficiently.}
        \label{figframe}
    \end{figure}The term indexed by $(n,m)$ is supported on the translate of ${\mathcal H}$ by $(nM, mN)$ and the crystallization condition implies that we can read off the taps of $h_{\text{eff}}[k,l]$ from the first term indexed by $(n,m) = (0,0)$. The terms corresponding to $(n,m) \ne (0,0)$ are sometimes also referred to as aliases of the first term. The delay and Doppler spread of the support set of the first term and the alias terms is the same, i.e., approximately $B\tau_s$ and $T \nu_s$ respectively (see paragraph after (\ref{eqn25})). 

    Fig.~\ref{figframe} shows a typical Zak-OTFS frame in the DD domain, formed of pilot, guard and data regions. The red dot, located at $(k_p, l_p)$, represents the transmitted pilot, and the pink ellipse represents the support set ${\mathcal S}$ of the received pilot. The yellow strip parallel to the Doppler axis represents the pilot region where no data pulsones are transmitted. It is dedicated to acquisition of the I/O relation. We transmit data pulsones from the green data regions, and we introduce blue guard regions to minimize interference between pilot and data. Data is not transmitted in the guard region, but the data signal received in the guard region is used for equalization. The overhead is the ratio of the number of carriers in the pilot and guard regions to the total number of carriers.  Note that as long as the crystallization condition holds, system overhead is independent of the channel Doppler spread $\nu_s < \nu_p$ since the length of the yellow strip (i.e., Doppler period $\nu_p$) is fixed. This is not the case with CP-OFDM where an increase in Doppler spread (reduction in coherence time) leads to an increase in pilot density.

\subsection{Predictability of I/O relation}
\label{subseciorel1}
For a given channel, the I/O relation of a modulation method is said to be predictable, if the known channel response to a  particular carrier waveform can be used to accurately predict/estimate the channel response to any other carrier waveform. From \cite{zakotfs1} we know that
the I/O relation with frequency domain modulation (FDM) is not predictable. However, the Zak-OTFS I/O relation is predictable when (\ref{eqn31}) is satisfied and the I/O relation can be acquired accurately and efficiently \cite{zakotfs2}. 

\subsection{To prevent ICI or to embrace ICI}
\begin{figure}
     \centering
        \vspace{-1mm}
        \includegraphics[width=7.12cm,height=4.3cm]{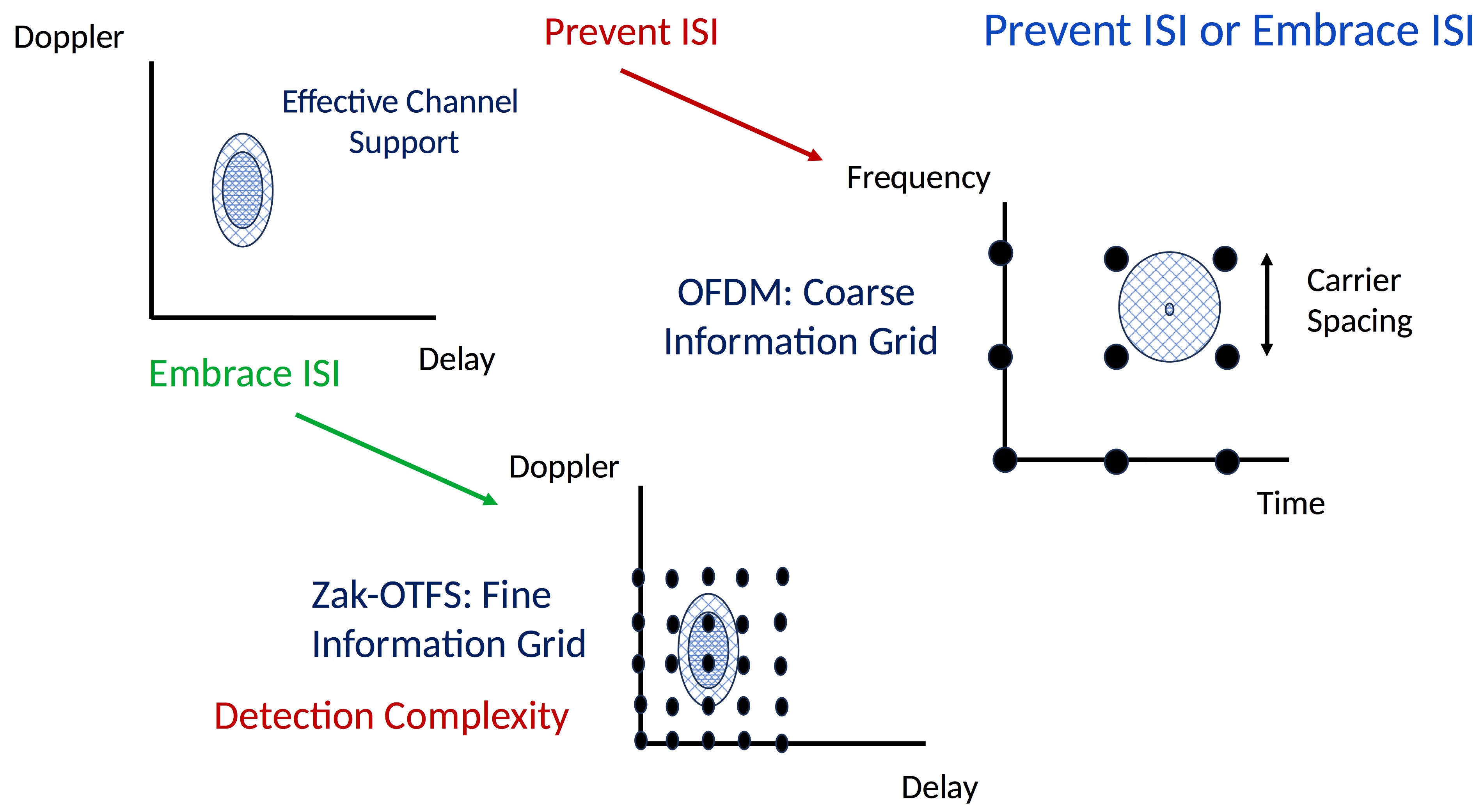}
        \vspace{-2mm}
        \caption{A fundamental architectural choice. CP-OFDM operates on a coarse information grid to avoid ISI and ICI. Zak-OTFS operates on a fine information grid with carrier waveforms designed to enable joint equalization.}
        \label{fig2}
        \vspace{-3mm}
    \end{figure}

This architectural choice is illustrated in Fig.~\ref{fig2},
and we begin by explaining how CP-OFDM limits CP overheads and avoids ICI. We choose the SCS $\Delta f$ to be significantly greater than the channel Doppler spread $\nu_s$, and we choose the symbol duration $T=1/\Delta f$ to be significantly greater than the CP duration $T_{cp}$. The information grid in CP-OFDM consists of points separated by $T$ seconds in time and by $\Delta f$ Hz in frequency, and each received sub-carrier is spread over approximately $(T+\tau_s)$ seconds and $(\Delta f + \nu_s)$ Hz. When $\Delta f \gg \nu_s$ and $T \gg \tau_s$, the CP-OFDM information grid is much coarser than that of the physical channel, and we are able to limit CP overhead and avoid ICI. This is illustrated in Fig.~\ref{fig2} which represents the delay and Doppler spread of the physical channel by a very small ellipse and represents the support of the effective channel (after pulse shaping and matched filtering) by a much larger outer ellipse. When channel delay and Doppler spreads are high, the drawback of this approach is that the pilot overhead becomes significant (see Section \ref{subsecpreventici}).

   We now turn to Zak-OTFS. The carrier spacing 1/B in delay is less than the channel delay spread, the carrier spacing 1/T is less than the channel Doppler spread, and we can simply read off the I/O relation from the response to a single pilot carrier waveform. The choice of a fine information grid embraces ICI. The carrier waveforms are designed to interfere with each other in the same way, making joint equalization possible. Note that joint equalization is possible in Zak-OTFS as its I/O relation can be acquired completely and efficiently whereas joint equalization of all OFDM sub-carriers is \emph{not possible} as it is challenging and difficult to acquire the CP-OFDM I/O relation completely and efficiently (see Section \ref{subseciorel2}). Equalization in Zak-OTFS is an active research area, and we have developed a joint equalization algorithm \cite{SRM2025} with the same linear order complexity ($O(BT)$) as that of per-carrier equalization in CP-OFDM. 
   

    \begin{figure}[h!]
    	\vspace{-7mm}
    	\begin{subfigure}[b]{0.5\textwidth}
    	\includegraphics[width=1.0\textwidth]{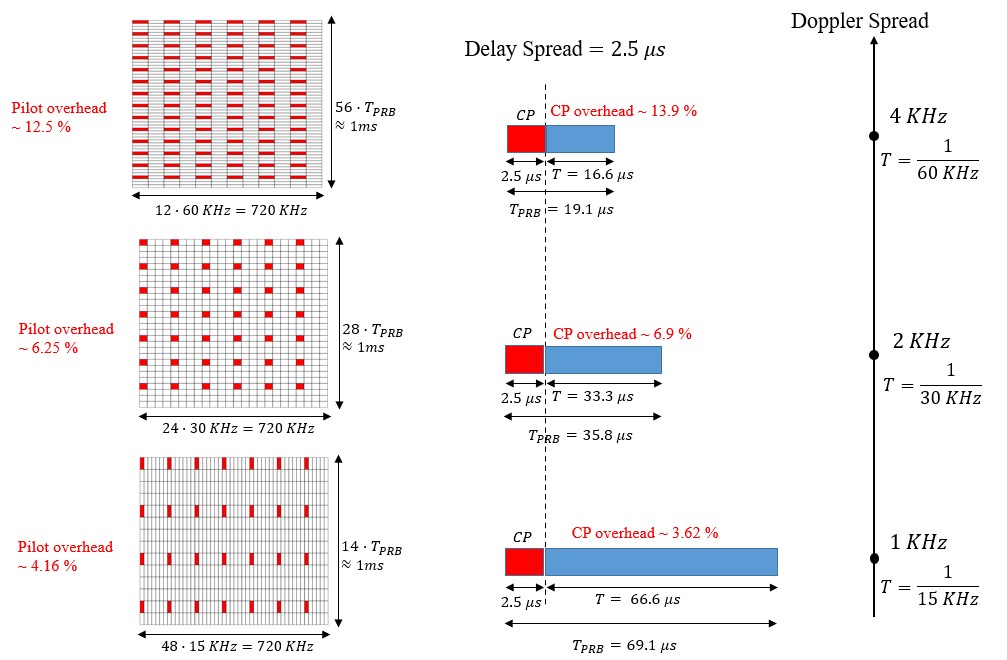}
    	\caption{}
    	\label{fig22}
    \end{subfigure}
    \begin{subfigure}[b]{0.5\textwidth}
	\includegraphics[width=1.0\textwidth]{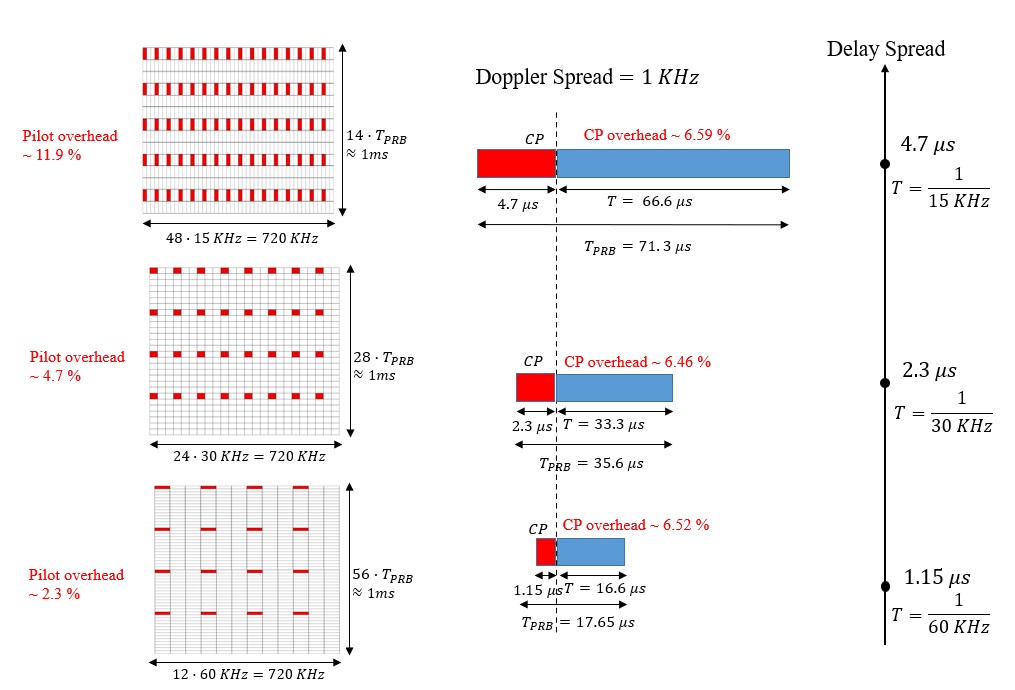}
	\caption{{}}
	\label{fig44}
	\end{subfigure}
	\vspace{-2mm}
	\caption{Effect of (a) channel Doppler spread and (b) delay spread on CP-OFDM overheads, for a CP-OFDM sub-frame of duration $1$ ms and $48$ sub-carriers.} 
 \label{fig94}
\end{figure}

\subsection{Cost of preventing ICI}
\label{subsecpreventici}
In CP-OFDM, SCS increases with channel Doppler spread to avoid ICI. This increases the CP overhead, because symbol duration decreases but CP duration remains the same, since it depends only on the channel delay spread.\footnote{\footnotesize{{In 3GPP 5G NR, numerology formats are specified in such a way that the CP overhead is the same. In other words, with every doubling in the SCS the CP duration is halved. The reason argued is that higher SCS would be used for higher carrier frequencies since the channel Doppler spread is higher for the same relative speed between transmitter and receiver and the required CP is smaller since the channel delay spread is smaller due to higher path loss.
 This standard argument for shrinking CP size with increasing SCS does not apply in our discussion here since we consider higher mobility without any increase in the carrier frequency.}}} As the Doppler spread increases, the channel coherence time decreases, so that pilots need to be transmitted more frequently in time, adding to the CP overhead. 

\uline{Example:} We assume a fixed delay spread of $1.15 \mu s$ and compare overheads in CP-OFDM and Zak-OTFS. Fig.~\ref{fig22} illustrates how the sum of pilot and CP overhead increases from $7.68\%$ to $26.4\%$ as the Doppler spread increases from $1$ kHz to $4$ kHz. With Zak-OTFS, we choose $\nu_p = 5$ kHz and $\tau_p = 1/\nu_p = 200 \mu s$ to satisfy the crystallization condition. The pilot and guard region overhead is only $2 \tau_s/\tau_p = 2 \times 2.5 / 200 = 2.5\%$. 

 Next we fix the channel Doppler spread and increase the channel delay spread. The symbol duration $T$ in CP-OFDM increases so the CP-overhead remains small. However, the channel coherence bandwidth decreases, and we need to increase pilot density in the frequency domain. Also, as $T$ increases and $\Delta f$ decreases, we sample the frequency domain more coarsely, which degrades the accuracy of channel estimates. 

\uline{Example:} We assume a fixed Doppler spread of $1$ kHz and compare overheads in CP-OFDM and Zak-OTFS. Fig.~\ref{fig44} illustrates how the sum of pilot and CP overhead increases from $8.82\%$ when the channel delay spread is $1.15 \mu s$ to $18.49\%$ when the delay spread is $4.7 \mu s$. With Zak-OTFS, we choose $\tau_p = 500 \mu s$ and $\nu_p = 1/\tau_p = 2$ kHz to satisfy the crystallization condition. The pilot and guard region overhead is only $2 \tau_s / \tau_p = 1.88\%$.

 The above comparisons assume that the pilot in Zak-OTFS is a single Zak-OTFS carrier separated from carriers used to transmit data by pilot and guard regions. We can avoid the overhead of pilot and guard regions entirely by employing a spread pilot with energy equally distributed across the Zak-OTFS carrier waveforms\cite{spreadpilotpaper}.

The above comparisons assume that all users share the same Doppler characteristics. However, CP-OFDM does not allow different users to have different SCS, and a single user subject to ICI causes carrier spacing to increase for all users. The user with the greatest Doppler spread determines the SCS for all users. Zak-OTFS is more flexible than CP-OFDM in that different users can have different delay and Doppler periods provided the crystallization condition is satisfied. We can also provision a single delay and Doppler period for all users and vary the pilot structure to accommodate different delay and Doppler spreads (see \cite{zakpilots} for more details).

\emph{In Zak-OTFS, accuracy and efficiency of I/O acquisition are almost \uline{decoupled} with the amount of channel delay and Doppler spread as long as the crystallization condition in (\ref{eqn31}) is satisfied. As explained in the following, Zak-OTFS pilot overhead is small and almost independent of the channel delay and Doppler spread, given that (\ref{eqn31}) is satisfied.} Let ${\Tilde \tau_s}$ and ${\Tilde \nu_s}$ denote the maximum possible channel delay and Doppler spread over all considered scenarios, i.e., $\tau_s \leq {\Tilde \tau_s}$ and $\nu_s \leq {\Tilde \nu_s}$. If ${\Tilde \tau_s} \, {\Tilde \nu_s} \ll \frac{1}{4}$, then $\nu_p = 2 {\Tilde \nu_s}$ and $\tau_p = 1/{\Tilde \nu_p} = 1/(2 {\Tilde \nu_s})$ satisfies the crystallization condition in (\ref{eqn31}) with small pilot overhead fraction $2 \tau_s/\tau_p \ll 1$, {\emph{for all}} channel scenarios where $\tau_s \leq {\Tilde \tau_s}$ and $\nu_s \leq {\Tilde \nu_s}$. This guarantees accurate and efficient acquisition of the I/O relation for all these doubly-spread channel scenarios. For the scenarios considered in this paper, ${\Tilde \tau_s} = 4.7 \mu s$ and ${\Tilde \nu_s} = 4$ kHz, which implies ${\Tilde \tau_s} \, {\Tilde \nu_s} = 0.0188 \ll 1/4$).   

   \begin{figure*}
   \vspace{-9mm}
     \centering
        \vspace{-1mm}\includegraphics[width=16.32cm,height=3.7cm]{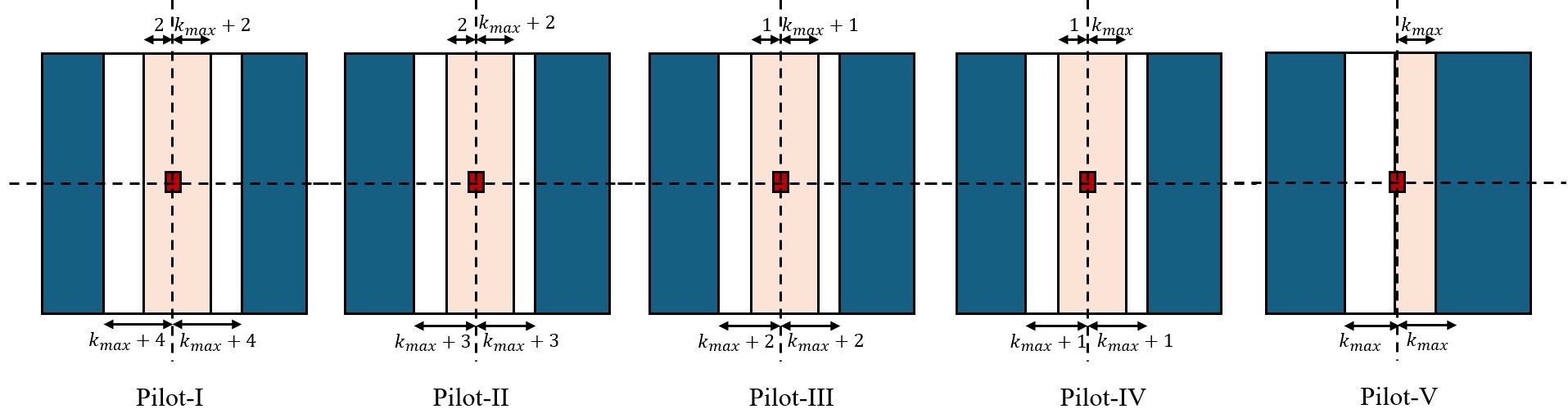}
        \vspace{-2mm}
        \caption{Zak-OTFS pilot allocations I to V.}
        \label{fig_pilotalloc}
        \vspace{-4mm}
    \end{figure*}
    \begin{table}[!t]
    \centering
    \caption{Power-delay profile of Veh-A channel model}
    \vspace{-2mm}
    \begin{tabular}{|c|c|c|c|c|c|c|}
         \hline
         Path index $i$ & 1 & 2 & 3 & 4 & 5 & 6 \\
         \hline
         Delay $\tau_i (\mu s)$ & 0 & 0.31 & 0.71 & 1.09 & 1.73 & 2.51 \\
         \hline
         Relative power (dB) & 0 & -1 & -9 & -10 & -15 & -20 \\
         \hline
    \end{tabular}
    \label{tab:veh_a}
    \vspace{-3mm}
\end{table}
\section{Choice of operating point}  
\label{secopoint}
The SCS $\Delta f$ determines the operating point for CP-OFDM, with $\Delta f$ increasing when the Doppler spread is high, and decreasing when the delay spread is high. The operating point for Zak-OTFS involves more choices and depends on (i) the delay and Doppler periods; (ii) the transmit pulse shaping filter; and (iii) the pilot and guard regions. We derive insight into how to choose the operating point through numerical simulations for the Veh-A channel model \cite{EVAITU} which consists of six channel paths and is representative of real propagation environments. The power delay profile is given in Table-\ref{tab:veh_a}, and we consider a Jakes Doppler spectrum, where the Doppler shift of the $i$-th path is given by $\nu_i = \nu_{max} \cos (\theta_i)$, where $\nu_{max}$ is the maximum Doppler shift, and $\theta_i$, $i=1,2,\cdots,6$ are i.i,d. random variables, uniformly distributed in $[0 \,,\, 2\pi)$.  The maximum possible Doppler spread is $\nu_s = 2 \nu_{max}$. We consider a single TF resource with $T=1$ ms and $B = 672$ kHz, and we fix the receiver SNR (data and pilot) to be $12$ dB. 

We optimize Zak-OTFS SE over the five different pilot allocations shown in Fig.~\ref{fig_pilotalloc}. Pilot allocation I minimizes interference between pilot and data at the cost of increasing the pilot/guard region overhead. We collect more pilot energy with a wider pilot/guard region, and that improves the accuracy of our estimate of the I/O relation, but we also collect more noise. The pilot/guard region overhead for pilot allocation n = 1, 2, 3, 4, 5 in  Fig.~\ref{fig_pilotalloc} is given by

{\small
\vspace{-4mm}
\begin{eqnarray}
\label{eqn34}
    \frac{(2k_{max} + 2(5 - n) + 1)}{M} & \hspace{-3mm} = & \hspace{-3mm} \nu_p \frac{(2\lceil B \tau_s \rceil + 2(5 - n) + 1)}{B}. \nonumber \\
\end{eqnarray}\normalsize}
We optimize Zak-OTFS SE over pilot to data ratio (PDR) values $-15, -10, -5, 0, 5, 10$, and $15$ dB. Increasing pilot power improves the accuracy of acquisition of the I/O relation, but the BER starts to degrade when interference from the pilot becomes more significant than noise. We optimize the pulse shaping filter, selecting from the Gauss, Gauss-sinc, and sinc filters described in \cite{Gaussinc}. We optimize over Doppler periods $\nu_p = 1, 2, 4, 6, 8, 12, 14$, and $24$ kHz. We perform joint equalization of all data pulsones using the Least Squares Minimum Residual (LSMR) joint equalizer \cite{lsmreq}. We also optimize the SE w.r.t. the modulation and coding scheme (MCS) combinations as specified by 3GPP \cite{3gppmcs1, 3gppmcs2}. Each MCS defined the QAM modulation order and the LDPC code rate, and we choose the highest MCS (highest information rate) for which the LDPC coded block error rate (BLER) is less than $0.1$. For any given operating point, the achieved SE is zero if
BLER is more than $0.1$ and otherwise it is given by $\eta = (1 - \text{BLER}) N_I / (B (T + \tau_s))$, where $N_I$ is the number of information bits transmitted in a Zak-OTFS frame.    

\subsection{Choice of Doppler period $\nu_p$}
   \begin{figure}
     \centering
        \vspace{-1mm}\includegraphics[width=8.32cm,height=5.7cm]{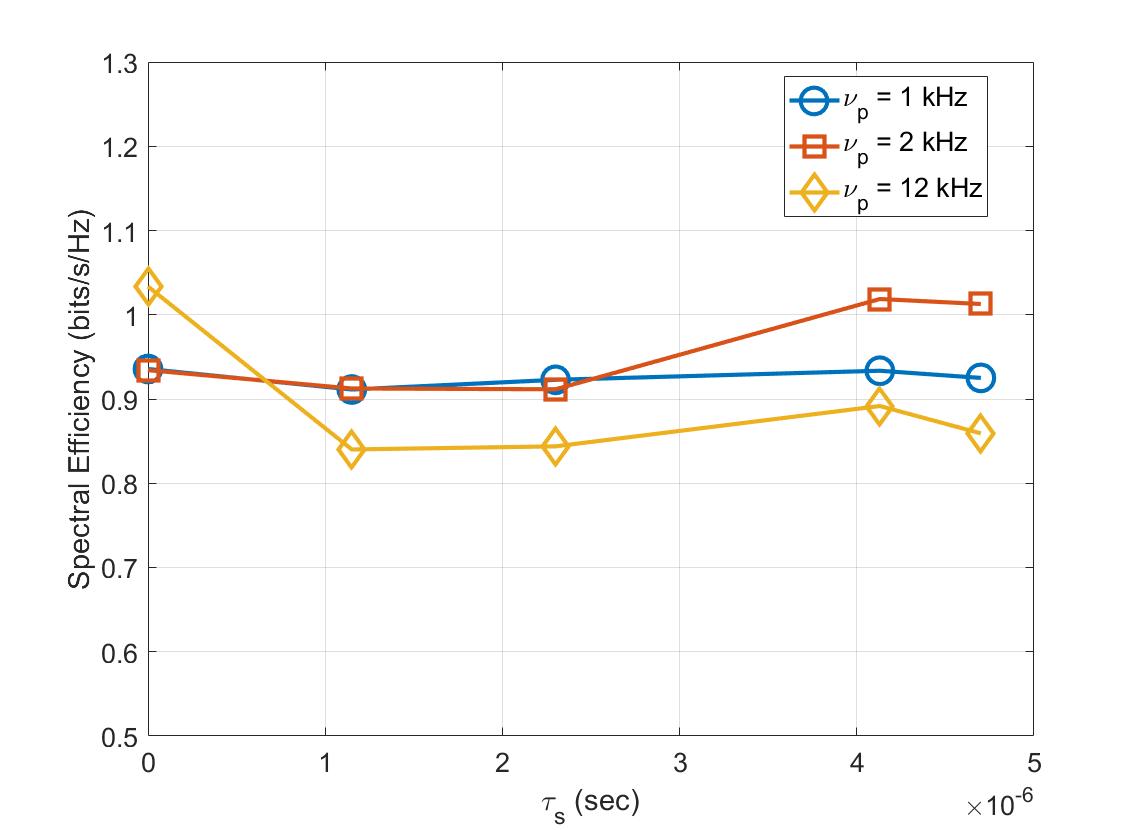}
        \vspace{-2mm}
        \caption{Zak-OTFS effective SE vs. delay spread ($\tau_s$). Low mobility ($\nu_{max} = 100$ Hz), SNR=$12$ dB, and Gauss-sinc pulse shaping.}
        \label{fig_op1}
        \vspace{-3mm}
    \end{figure}
    On the one hand, when we increase $\nu_p$ we improve the accuracy of acquisition of the I/O relation by reducing Doppler aliasing in the pilot region.\footnote{\footnotesize{By aliasing we mean that the first term and alias terms in (\ref{eqn33}) have overlapping support which degrades the accuracy of the acquired I/O relation.}} On the other hand, we increase the overhead of the pilot/guard region given by (\ref{eqn34}). We optimize SE by balancing these two effects, and the Doppler period $\nu_p$ that optimizes SE will depend on the type of wireless channel. In this Section, we investigate channels with low Doppler and high delay spreads, also channels with moderate to high Doppler spreads.

    We begin with a low-mobility scenario where $\nu_{max} = 100$ Hz. Fig.~\ref{fig_op1} plots optimized SE as a function of increasing delay spread  $\tau_s = 0, 1.15, 2.3, 4.13, 4.7 \mu s$ for different $\nu_p = 1,2, 12$ kHz (for a fixed total SNR of $12$ dB and fixed $\nu_p$, SE is optimized w.r.t. the other parameters). Note that the crystallization condition is satisfied for all these delay spreads. Table-\ref{tab:veh_a} specifies a Veh-A channel model with delay spread $2.5 \mu s$, and we simply scale all path delays so that the sixth and longest path has delay $\tau_s$.
    
The blue curve in Fig.~\ref{fig_op1} shows optimized SE for the minimum possible Doppler period $\nu_p = 1/T = 1$ kHz and $N = T \nu_p = 1$. Since $N$ is small, there is aliasing along the Doppler axis due to the spread of the pulse shaping filter. \emph{Doppler domain aliasing limits performance and increasing the size of the pilot / guard region by varying the pilot allocation does not significantly improve SE.}

The red curve in Fig.~\ref{fig_op1} shows optimized SE for Doppler period $\nu_p = 2$ kHz, where $N=2$ and $M=336$. There is less aliasing along the Doppler axis, and at higher delay spreads $\tau_s$ we are able to make use of MCS with higher rates by increasing the width of the pilot region. For example, MCS-$8$ is achieved for $\tau_s = 4.13, 4.7 \mu s$, and only MCS-$7$ for the other delay spreads. Pilot allocation IV is optimal for both $\tau_s = 4.13\mu s$ and $\tau_s = 4.7 \mu s$ and the corresponding overheads are $9/336 = 2.7\%$ and $11/336 = 3.3\%$ respectively.


The yellow curve in Fig.~\ref{fig_op1} shows optimized SE for a much higher Doppler period $\nu_p = 12$ kHz, where $N=12$ and $M=56$. Since $\nu_{max} = 100$ Hz is very much smaller than $\nu_p$, there is essentially no Doppler domain aliasing. When there is no delay spread, there is no interference between pilot and data, we are able to make use of higher rate MCS and thereby improve upon spectral efficiencies that can be achieved for $\nu_p = 1$ and $2$ kHz. As the delay spread increases, so does the pilot/guard region overhead. For example, when the delay spread $\tau_s = 4.7 \mu s$, we choose pilot allocation IV, and the resulting overhead $11/56$ is about $20\%$, which is much higher than the $3.3\%$ overhead achieved with $\nu_p = 2$ kHz. The large overhead leads to a SE inferior to that which can be achieved for $\nu_p = 1$ and $2$ kHz.

\emph{In low mobility scenarios ($\nu_{max} < 1/2T$) and moderate to high delay spreads ($\tau_s > 1/B$), the optimal Doppler period $\nu_p$ should be large enough to provide good estimates of the I/O relation ($\nu_p \geq 1/T$) without creating a pilot/guard region that is needlessly large.}

\begin{figure}
     \centering
        \vspace{-5mm}\includegraphics[width=8.32cm,height=5.7cm]{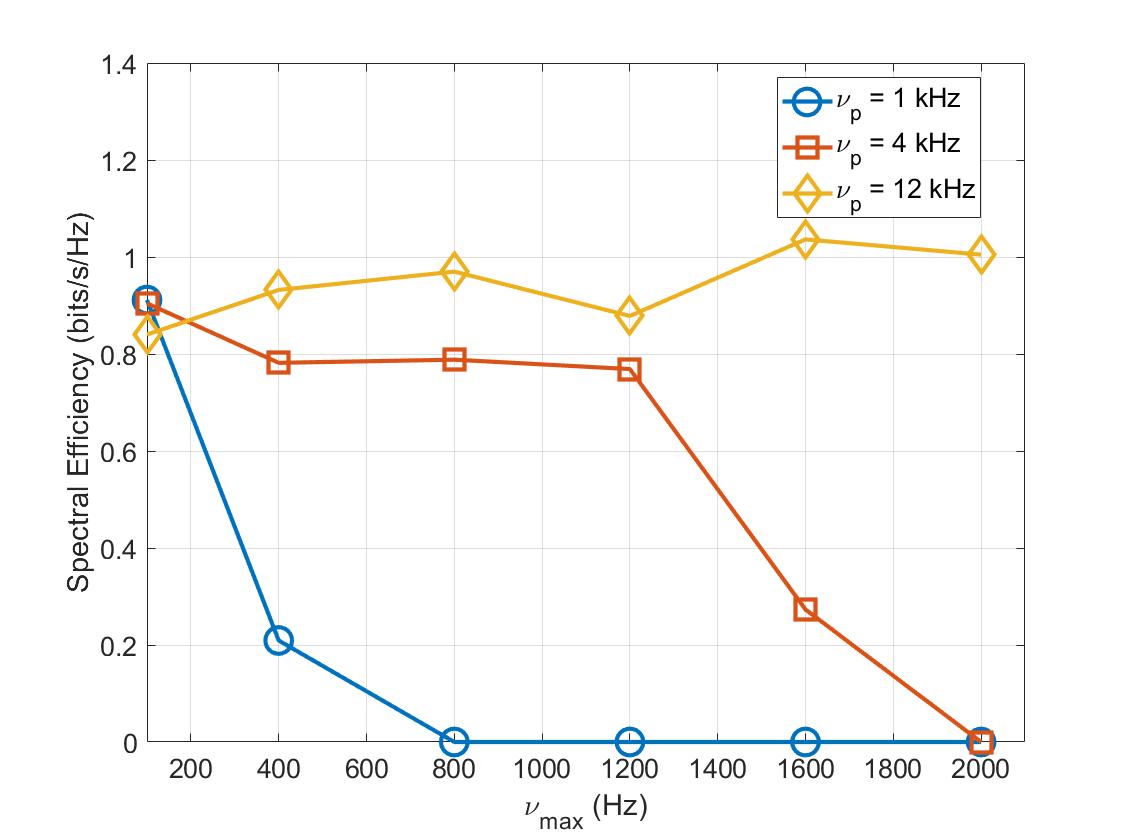}
        \vspace{-2mm}
        \caption{Zak-OTFS effective SE vs. max. Doppler shift ($\nu_{max} = \nu_s/2$). Small cell ($\tau_s = 1.15 \mu s$), SNR=$12$ dB, and Gauss-sinc pulse shaping.}
        \label{fig_op2}
        \vspace{-3mm}
    \end{figure}

Fig.~\ref{fig_op2} illustrates the variation in optimized SE as a function of increasing $\nu_{max} = 100, 400, 800, 1200, 1600, 2000$ Hz, for a fixed delay spread $\tau_s = 1.15 \mu s$. SE drops dramatically when the crystallization condition is not satisfied. When $\nu_p = 1$ kHz, the crystallization condition is not satisfied for $\nu_{max} \geq 500$ Hz, and the BLER exceeds $0.1$ for all MCS. When $\nu_p = 4$ kHz, SE is zero for $\nu_{max} \geq 2$ kHz.

\emph{In moderate to high mobility scenarios ($\nu_{max} > 1/2T$), the optimal Doppler period $\nu_p$ should be large enough to satisfy the crystallization conditions. Simulations suggest that SE is good when we choose $\nu_p$ to be about $4 \nu_{max}$ provided the pilot overhead is small ($2 \tau_s \ll \tau_p$, or equivalently $8 \tau_s \nu_{max} \ll 1$).}  

In this paper, we have considered scenarios where $\tau_s \leq 4.7 \mu s$ and $\nu_{max} \leq 2$ kHz, so that $8 \tau_s \nu_{max} \leq 0.0752 \ll 1$. The above conditions can be satisfied, and when they are satisfied, SE is good.  

\subsection{Choice of pulse shaping filter}
We consider three pulse shaping filters with different characteristics. The Gauss filter is ideal for estimating the I/O relation since it leaks very little energy outside the main lobe. However, the main lobe is wide, and this degrades equalizer performance because it creates significant interference between neighboring pulsones. The sinc filter has a narrow main lobe, so equalizer performance is better than the Gauss filter. However, it leaks energy outside the main lobe, and this degrades estimation of the I/O relation. The Gauss-sinc filter is the product of a Gaussian pulse and a sinc pulse and inherits the positive attributes of both filters.

\begin{figure}
     \centering
        \vspace{-5mm}\includegraphics[width=8.32cm,height=5.7cm]{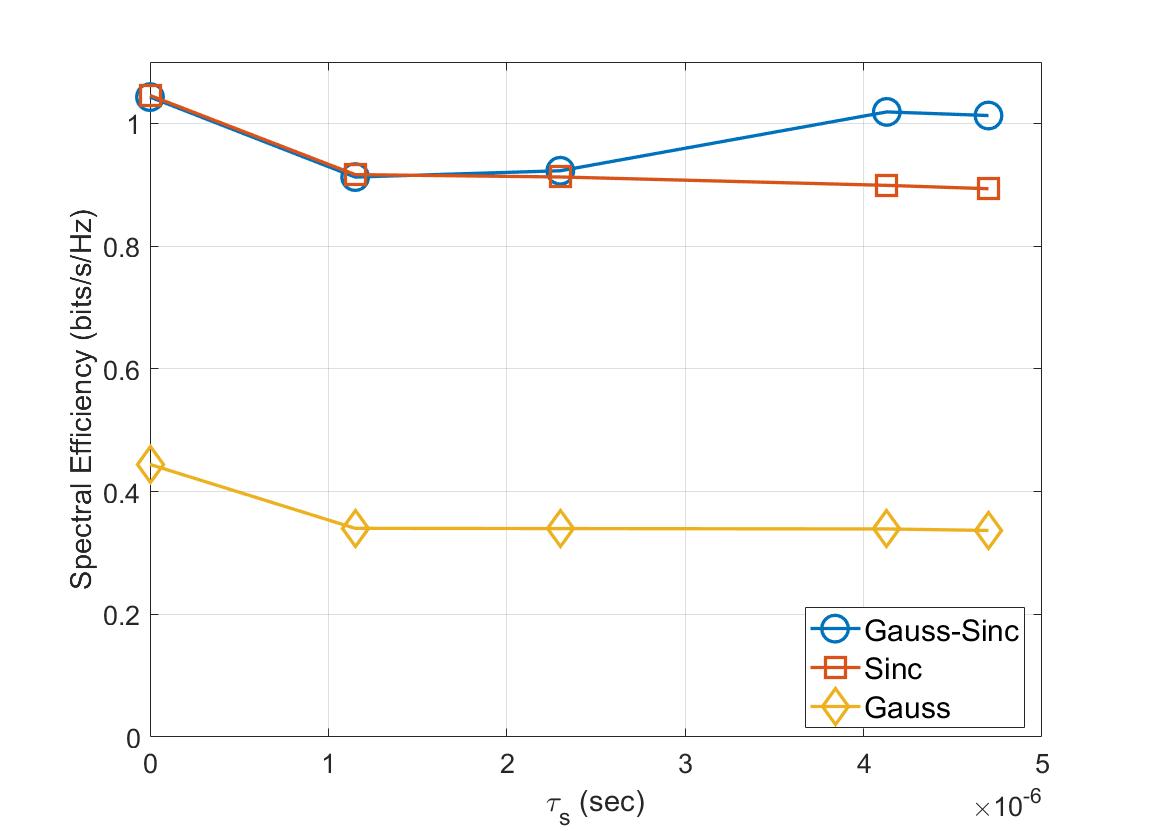}
        \vspace{-2mm}
        \caption{Zak-OTFS effective SE vs. channel delay spread ($\tau_{s}$) for different pulse shaping filters. Low mobility ($\nu_{max} = 100$ Hz) and SNR=$12$ dB.}
        \label{fig_op4}
        \vspace{-3mm}
    \end{figure}

Fig.~\ref{fig_op4} illustrates the variation in optimized SE as a function of increasing delay spread $\tau_s$ for a low mobility scenario ($\nu_{max} = 100$ Hz). The Gauss filter exhibits much poorer SE than the sinc and Gauss-sinc filters. For low to moderate delay spreads, the sinc and Gauss-sinc filters exhibit very similar SE because interference between pilot and data is not significant.
Interference becomes more significant at higher delay spreads ($\tau_s = 4.7 \mu s$), and because the Gauss-sinc filter is better localized, we are able to improve SE by accessing higher MCS. 

\begin{figure}
     \centering
        \vspace{-1mm}\includegraphics[width=8.32cm,height=5.7cm]{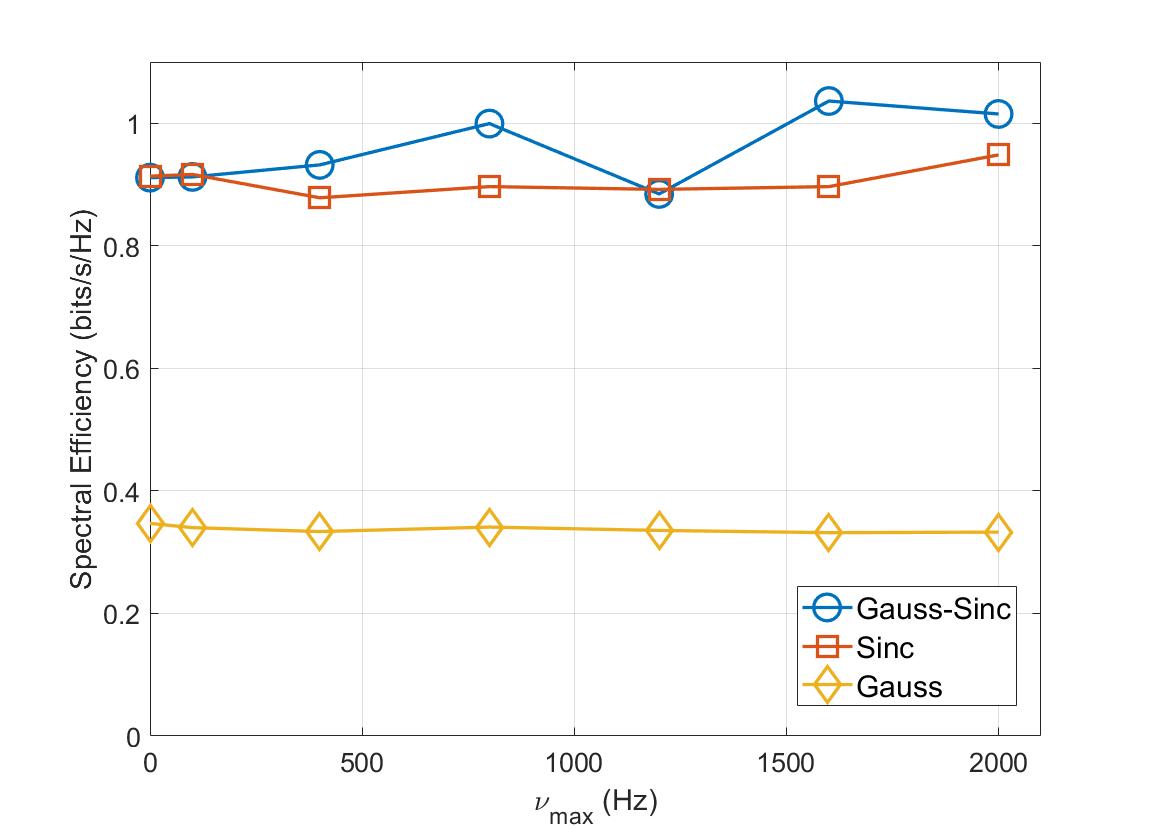}
        \vspace{-2mm}
        \caption{Zak-OTFS effective SE vs. max. Doppler shift ($\nu_{max}$) for different pulse shaping filters. Small cell ($\tau_s = 1.15 \mu s$) and SNR=$12$ dB.}
        \label{fig_op5}
        \vspace{-3mm}
    \end{figure}

    Fig.~\ref{fig_op5} plots optimized SE as a function of increasing $\nu_{max}$ for a fixed delay spread $\tau_s = 1.15 \mu s$. At low mobility, Doppler aliasing is insignificant, and the sinc and Gauss-sinc filters exhibit similar performance. As mobility increases, the Doppler period $\nu_p$ increases to control Doppler domain aliasing. As the pilot region is a strip along the Doppler axis with data pulsones on both sides, increasing $\nu_p$ introduces interference between data and pilot, making localization more important. This is why the performance of Gauss-sinc filter is superior. 

\emph{The Gauss-sinc filter exhibits superior SE across all scenarios because it balances accurate estimation of the I/O relation (governed by leakage outside the main lobe) with control of interference between data and pilot (governed by the width of the main lobe).}

\subsection{Choice of pilot allocation}
\begin{figure}
\hspace{-2mm}
\begin{subfigure}[t]{0.45\linewidth}
        \includegraphics[width=1.2\linewidth]{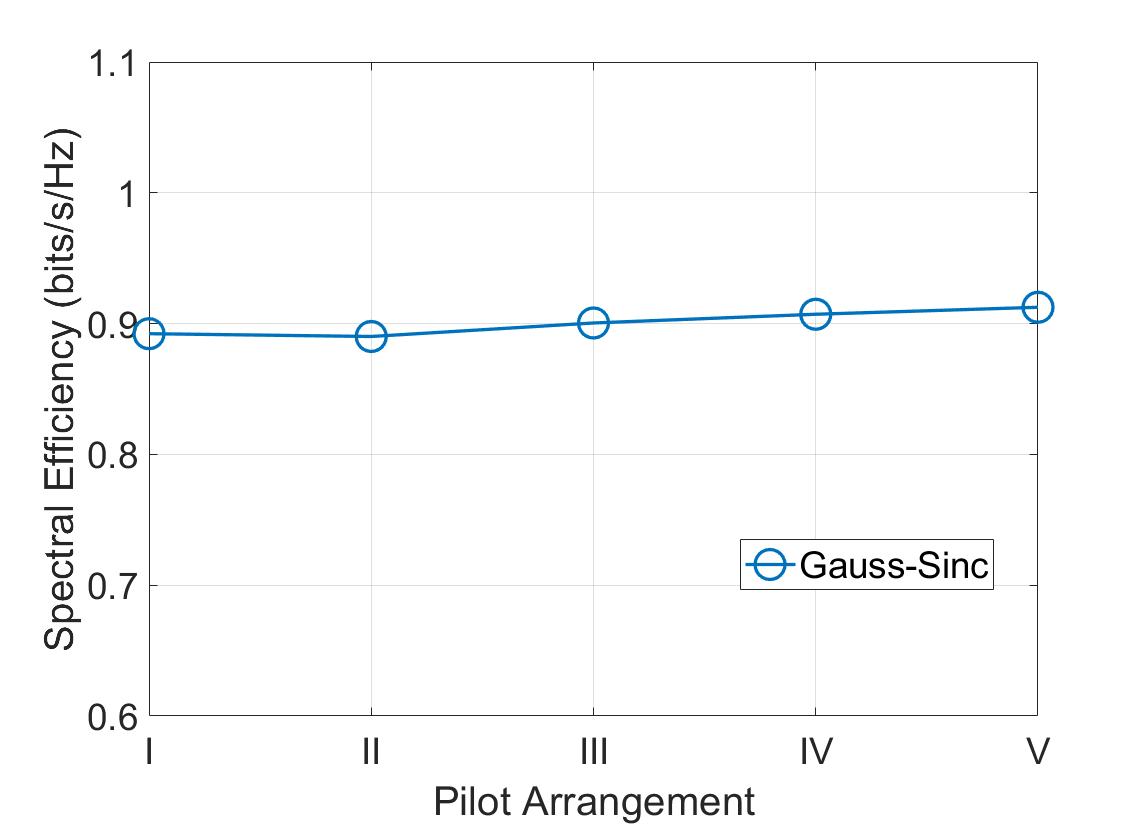}
        \vspace{-4mm}
        \caption{$\tau_s = 1.15 \mu s$.}
         \label{fig6:subfigA}
    \end{subfigure}\hspace{6mm} 
    \begin{subfigure}[t]{0.45\linewidth}
        \includegraphics[width=1.2\linewidth]{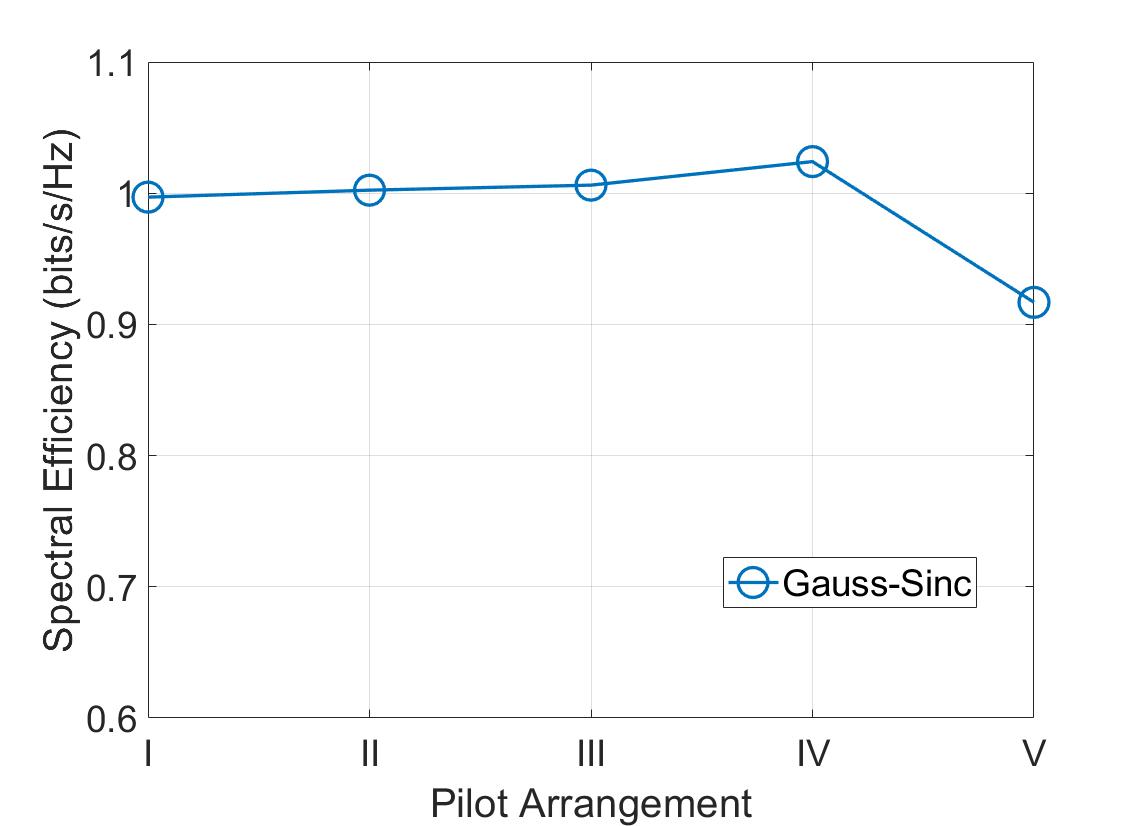}
        \vspace{-4mm}
        \caption{$\tau_s = 4.7 \, \mu s$.}
         \label{fig7:subfigB}
    \end{subfigure}
    \caption{Zak-OTFS effective SE vs. pilot allocation. Low mobility ($\nu_{max} = 100 $ Hz) with Doppler period $\nu_p = 2$ kHz.}
    \label{fig68}
\end{figure}

The width of the pilot/guard region decreases from pilot allocation I to pilot allocation V. A wider region increases overhead but might make it possible to access higher MCS by reducing interference between pilot and data. The effect of choosing the pilot allocation on SE will depend on the wireless scenario. Fig.~\ref{fig68} illustrates a low mobility scenario ($\nu_{max} = 100$Hz). We fix the Doppler period $\nu_p = 2$ kHz and we plot SE for the five pilot allocations. Fig.~\ref{fig79} replicates this simulation for a high mobility scenario ($\nu_{max} = 2$ kHz), where we fix the Doppler period $\nu_p = 12$ kHz. We work with a Gauss-sinc pulse shaping filter in both sets of simulations.


The delay spread $\tau_s$ governs the interference between pilot and data. In low mobility scenarios with small delay spread ($\tau_s = 1.15 \mu s$), $\nu_p$ is small and $\tau_p$ is large, and it follows from (\ref{eqn34}) that the pilot/guard region overhead is small. Due to small delay spread data-pilot interference is not significant and therefore SE is almost invariant of the pilot allocation, as can be observed in Fig.~\ref{fig6:subfigA}. When the delay spread is high ($\tau_s = 4.7 \mu s$), as we move from pilot allocation IV to V, the interference between pilot and data renders some MCS inaccessible and Fig.~\ref{fig7:subfigB} shows a decrease in SE.

In high mobility scenarios, $\nu_p$ is sufficiently large and $\tau_p$ is sufficiently small. In Fig.~\ref{fig68} as we move from pilot allocation I to II, SE reduces because interference between pilot and data renders higher MCS inaccessible. However, from allocation II to V, SE improves monotonically since reduction in pilot overhead is the dominant effect.  

\emph{In low mobility and low delay spread scenarios ($\nu_{max} < 1/(2T)$, $\tau_s < 1/B$), SE is almost independent of the pilot allocation (pilot/ guard overhead is small and interference between pilot and data is not significant). In low mobility and high delay spread scenarios ($\nu_{max} < 1/(2T)$, $\tau_s > 1/B$), when the pilot allocation is sufficiently narrow, SE reduces because interference between pilot and data renders higher MCS inaccessible. In high mobility scenarios  ($\nu_{max} > 1/(2T)$, the delay period $\tau_p$ is small, SE is determined by the pilot/guard region overhead, and pilot allocation V is optimal.}

\begin{figure}
\label{fig79}
\hspace{-2mm}
\begin{subfigure}[t]{0.45\linewidth}
        \includegraphics[width=1.2\linewidth]{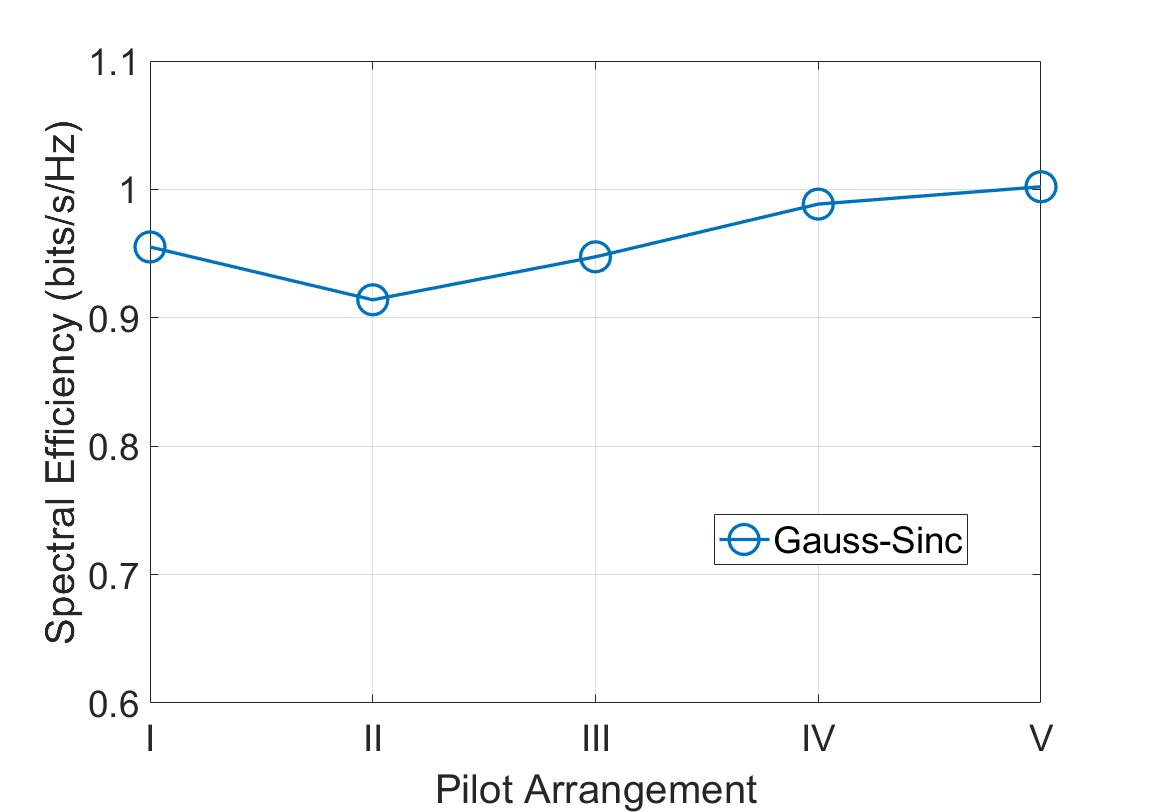}
        \vspace{-4mm}
        \label{fig8:subfigA}
        \caption{$\tau_s = 1.15 \, \mu s$.}
    \end{subfigure}\hspace{6mm} 
    \begin{subfigure}[t]{0.45\linewidth}
        \includegraphics[width=1.2\linewidth]{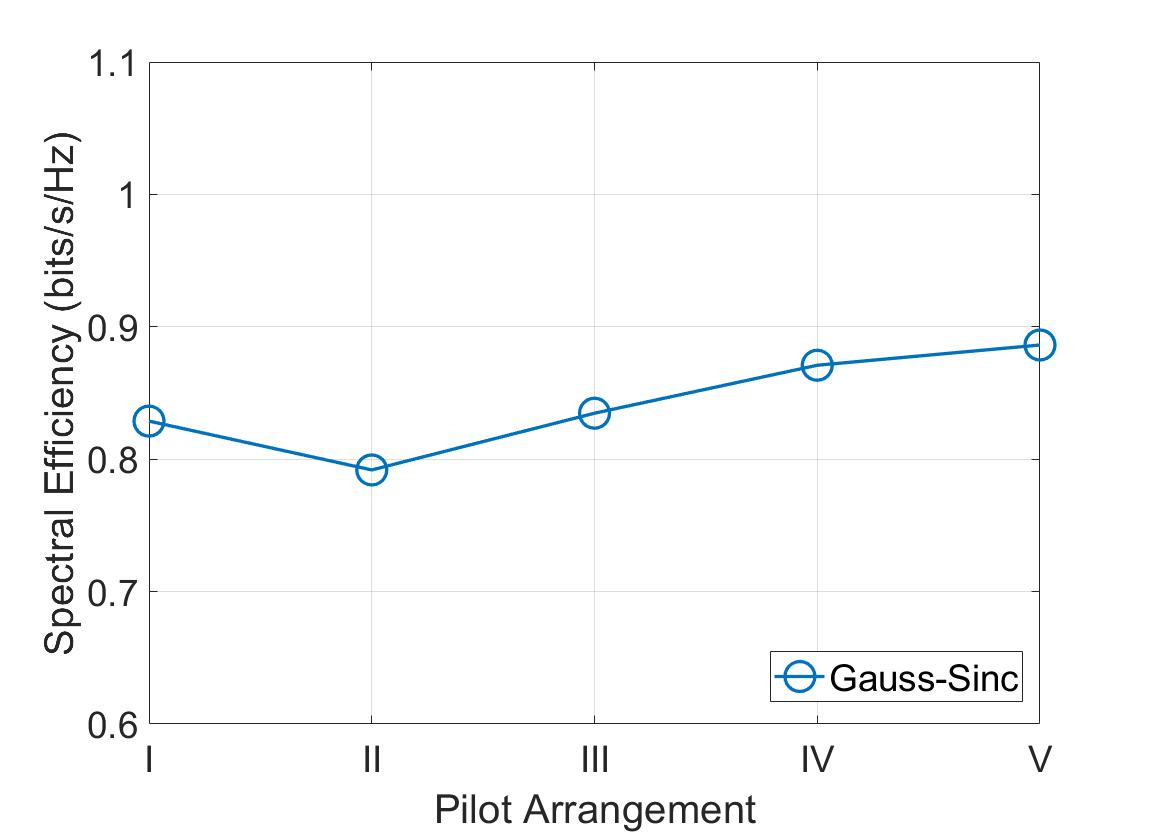}
        \vspace{-4mm}
        \label{fig9:subfigB}
        \caption{$\tau_s = 4.7 \, \mu s$.}
    \end{subfigure}
    \caption{Zak-OTFS effective SE vs. pilot allocation. High mobility ($\nu_{max} = 2$ kHz) with Doppler period $\nu_p = 12$ kHz.}
    \label{fig79}
\end{figure}

\subsection{Choosing the ratio of pilot power to data power (PDR)}
\begin{figure}
\hspace{-3mm}
\begin{subfigure}[t]{0.45\linewidth}
        \includegraphics[width=1.25\linewidth]{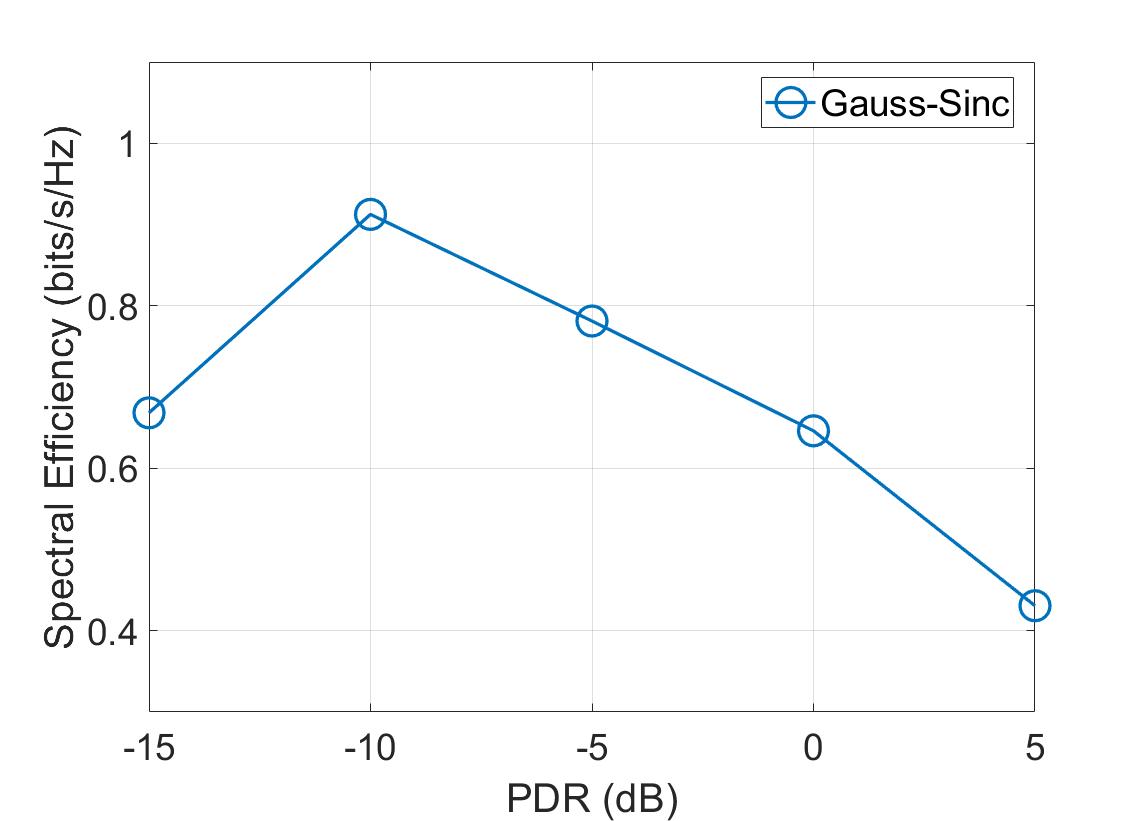}
        \vspace{-4mm}
        \label{fig11:subfigA}
        \caption{$\tau_s = 1.15 \, \mu s$.}
    \end{subfigure}\hspace{6mm} 
    \begin{subfigure}[t]{0.45\linewidth}
        \includegraphics[width=1.25\linewidth]{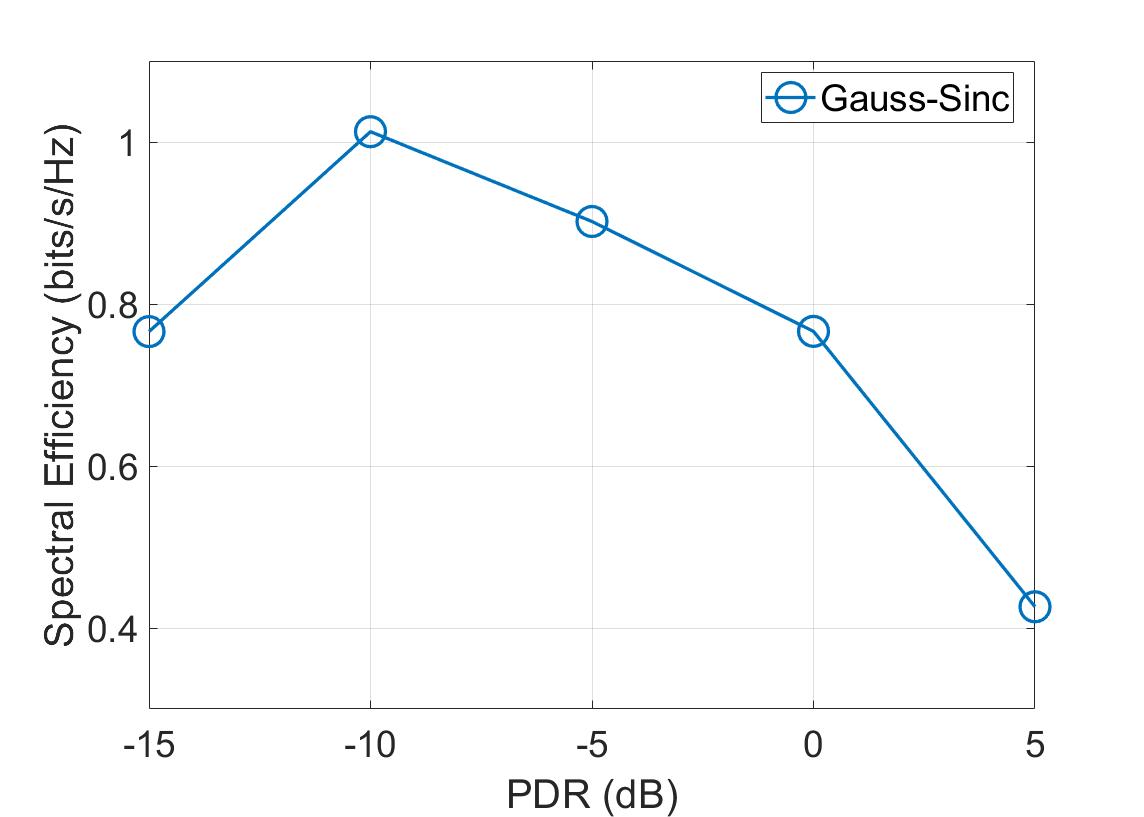}
        \vspace{-4mm}
        \label{fig12:subfigB}
        \caption{$\tau_s = 4.7 \, \mu s$.}
    \end{subfigure}
    \caption{Zak-OTFS effective SE vs. pilot to data power ratio (PDR). Low mobility ($\nu_{max} = 100$ Hz) with Doppler period $\nu_p = 2$ kHz.}
    \label{fig81}
\end{figure}

On the one hand a high PDR provides a more accurate estimate of the I/O relation and might make higher MCS accessible. On the other hand, it means more interference from pilot to data and a lower SNR for the data signal. There is therefore an optimal PDR.
We first consider low mobility ($\nu_{max} = 100$ Hz) using a Gauss-sinc filter for a fixed SNR $= 12$ dB, and for delay spreads $\tau_s = 1.15 \mu s$ and $\tau_s = 4.7 \mu s$. Fig.~\ref{fig81} plots SE (optimized over the remaining parameters) as a function of PDR. Since $\nu_p = 2$ kHz $\gg \nu_s$, there is almost no Doppler domain aliasing. Since $\tau_p \gg \tau_s$, the overhead of the widest pilot allocation is small, interference from pilot to data is not significant, and the accuracy of I/O estimation is limited only by the noise power. The power of the received pilot is localized to the pilot region, which is about $2 \tau_s/\tau_p$ of the total resource, whereas noise is spread equally across all $MN$ pulsones. Therefore, the effective pilot to noise ratio (PNR) is about $\text{PDR} \times \text{SNR} \times \frac{\tau_p}{2\tau_s}$. For a PDR $= -10$ dB, the effective PNR is about $19.3$ dB for $\tau_s = 4.7 \mu s$. This is sufficient to accurately estimate the I/O relation since the weakest path in the power delay profile listed in Table-\ref{tab:veh_a} is about $20$ dB below the strongest path. Instead, if PDR were to be $-15$ dB, the acquired I/O relation is not as accurate since the effective PNR is around $14.3$ dB, resulting in significant degradation in the achieved SE (see Fig.~\ref{fig81}). 

\begin{figure}
\hspace{-3mm}
\begin{subfigure}[t]{0.45\linewidth}
        \includegraphics[width=1.25\linewidth]{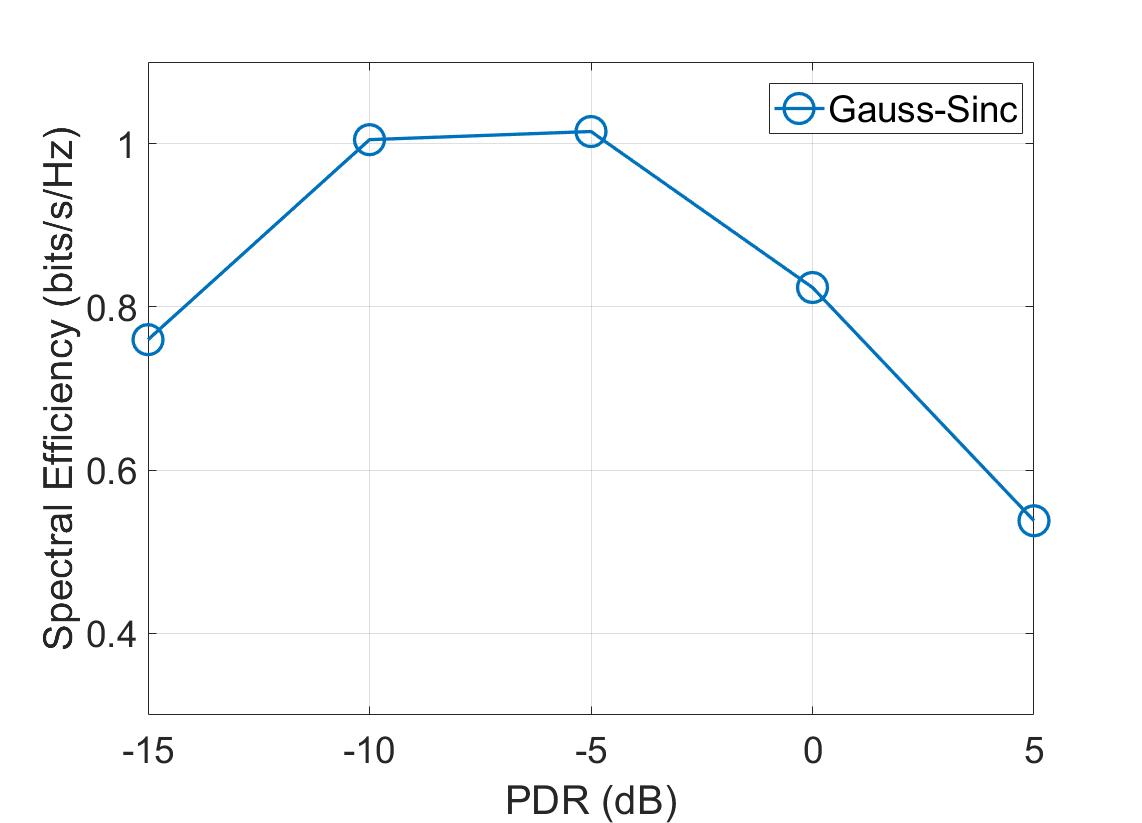}
        \vspace{-4mm}
        \label{fig13:subfigA}
        \caption{$\tau_s = 1.15 \, \mu s$.}
    \end{subfigure}\hspace{6mm} 
    \begin{subfigure}[t]{0.45\linewidth}
        \includegraphics[width=1.25\linewidth]{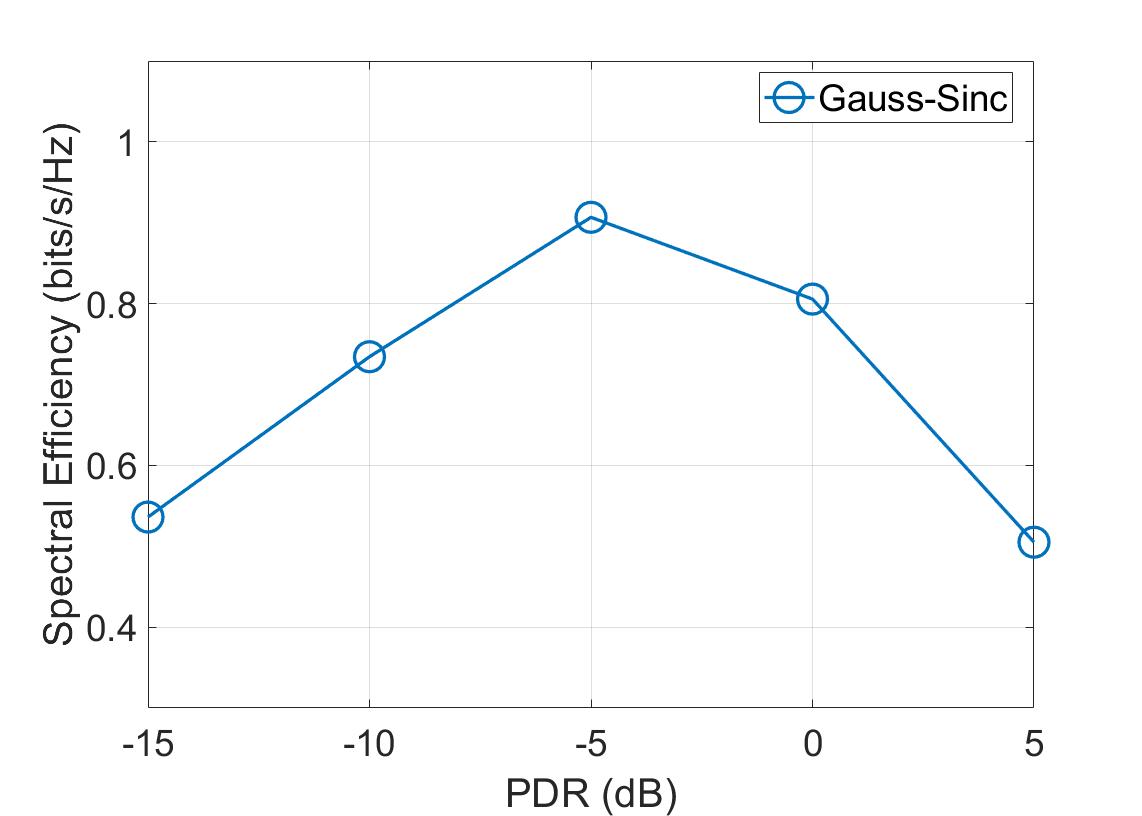}
        \vspace{-4mm}
        \label{fig14:subfigB}
        \caption{$\tau_s = 4.7 \, \mu s$.}
    \end{subfigure}
    \caption{Zak-OTFS SE vs. pilot to data power ratio (PDR). High mobility ($\nu_{max} = 2$ kHz) with Doppler period $\nu_p = 12$ kHz.}
    \label{fig83}
\end{figure}

We now carry out the same analysis for high mobility ($\nu_{max} = 2$ kHz), increasing the Doppler period $\nu_p$ to $12$ kHz so that the crystallization conditions are satisfied. This decreases both the delay period $\tau_p$ and the effective PNR $\text{PDR} \times \text{SNR} \times \frac{\tau_p}{2\tau_s}$, hence we need to increase the PDR to estimate the I/O relation. This explains why the optimal PDR in Fig.~\ref{fig83} is $-5$ dB, which is $5$ dB more than the optimal value for the low mobility scenario.

       \begin{figure}
        \vspace{-8mm}\includegraphics[width=9.12cm,height=6.7cm]{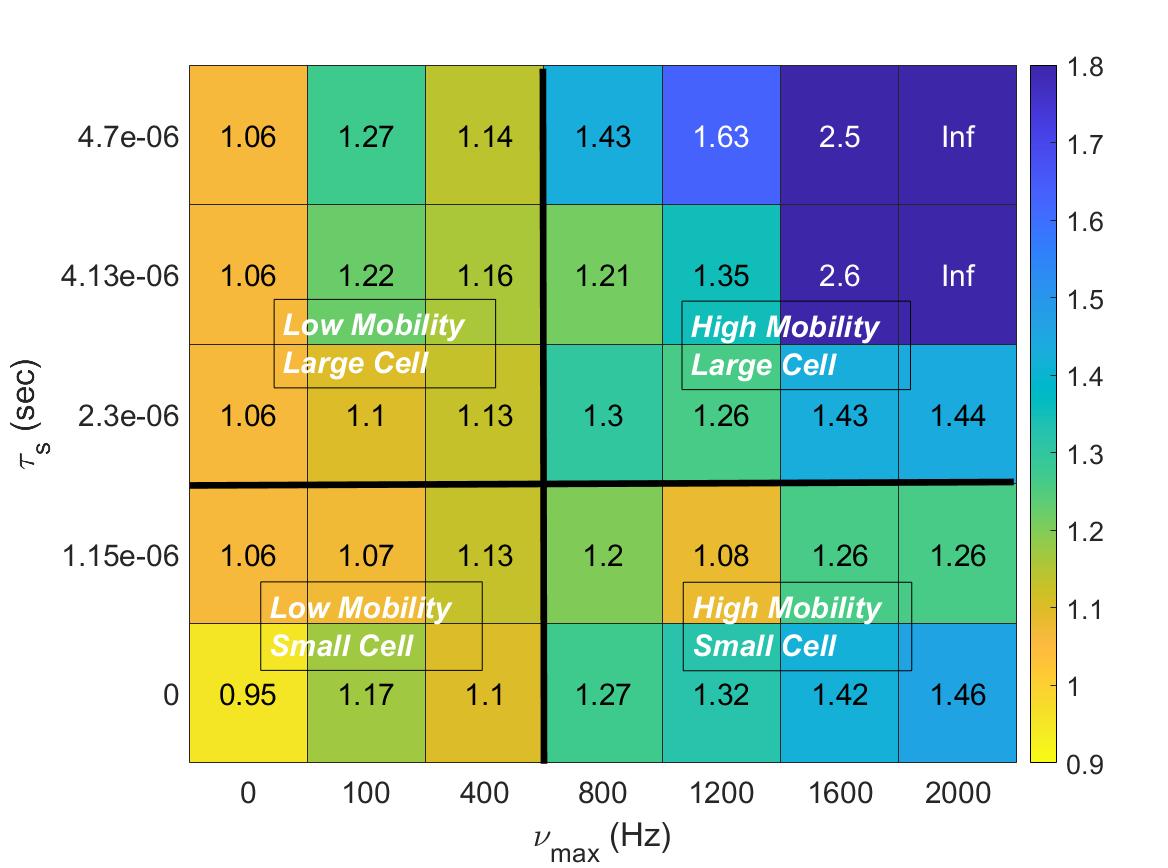}
        \vspace{-3mm}
        \caption{Heatmap showing the ratio of SE achievable with Zak-OTFS to that achievable with CP-OFDM for $35$ Doppler-delay combinations $(\nu_{max}, \tau_s)$. Target block error rate (BLER) = $0.1$ and SNR=$12$ dB.}
        \label{fig_op6}
        \vspace{-4mm}
    \end{figure}
\section{Comparing Spectral Efficiency of Zak-OTFS and CP-OFDM}
\label{sec7}
We fix the BLER $= 0.1$,we fix SNR $= 12$ dB, and we calculate the ratio SE(Zak-OTFS)/SE(CP-OFDM) of the optimized spectral efficiencies achieved with Zak-OTFS and with CP-OFDM. This ratio (relative SE) depends on mobility (low/high) and cell size (small/large), and we optimize SE for $35$ combinations ($\nu_{max}, \tau_s$), where $\nu_{max} = 0, 100, 400, 800, 1200, 1600, 2000$ Hz, and $\tau_s = 0, 1.15, 2.3, 4.13, 4.17 \mu s$. The $35$ combinations correspond to the $35$ cells shown in Fig.~\ref{fig_op6}, and the number in the cell is the relative spectral efficiency SE(Zak-OTFS)/SE(CP-OFDM). 

The physical resource for CP-OFDM is a slot of $14$ symbols and $4$ PRBs ($48$ sub-carriers), and we optimize SE over the possible SCSs ($15$, $30$, and $60$ kHz). When the delay spread $\tau_s= 4.13 \mu s$, the SCS can be $15$ kHz or $60$ kHz (extended CP with $12$ symbols).  When $\tau_s = 4.7 \mu s$, SCS is $15$ kHz, when $\tau_s = 2.3 \mu s$, SCS can be $15$ or $30$ kHz, and when $\tau_s = 0$ or $1.15 \mu s$, SCS can be $15$, $30$ or $60$ kHz. For both CP-OFDM and Zak-OTFS, we optimize SE over the modulation and coding schemes (MCS) specified in the 3GPP 5G NR standard \cite{3gppmcs1, 3gppmcs2}. For CP-OFDM we optimize over all possible SCSs, over all MCS, over Type-A DMRS pilot allocations, and over DMRS power boosts of $-6, -4, -2, 0, 2, 4$, and $6$ dB relative to the power on each data sub-carrier. For CP-OFDM, we perform per-sub-carrier MMSE equalization. For both CP-OFDM and Zak-OTFS, we include data and pilots in the total signal power, we fix the total SNR to be $12$ dB, and we select the combination that maximizes SE.  For Zak-OTFS, the SE is given by $(1 - \text{BLER}) N_I / (672\times 10^3  \times (10^{-3} + \tau_s))$ and for CP-OFDM it is given by $(1 - \text{BLER}) N_I / 720$. 

From the heat-map plot of the ratio of the optimized Zak-OTFS SE to the optimized CP-OFDM SE in Fig.~\ref{fig_op6} it is clear that while Zak-OTFS and CP-OFDM achieve similar SE 
for low-mobility small-cell scenarios, the performance improvement with Zak-OTFS is significant when either the delay-spread or the Doppler-spread is high. We now investigate performance in each of the four quadrants shown in Fig.~\ref{fig_op6}. 

    \begin{table}[!t]
    \vspace{-7mm}
    \centering
    \caption{Optimal param. vs $\nu_{max}$ (in Hz), fixed $\tau_s = 1.15 \mu s$. (Zak-OTFS, CP-OFDM) for each parameter.}
   \scriptsize{
    \begin{tabular}{|c|c|c|c|c|}
         \hline
         $\nu_{max}$ (in Hz) &  $100$ & $400$ & $800$ &  $1600$ \\
         \hline
         MCS &  $7,7$ & $8,7$ & $8,7$ &   $9,7$ \\
         \hline
         Overhead ($\%$) & $1.0, 8.8$ & $12.6, 11$ & $6.05, 11$ & $12.6, 13.2$ \\
         \hline
         BLER &  $0.1, 0.09$ & $0.09, 0.1$ & $0.09, 0.09$ & $0.1, 0.07$\\
         \hline
         SE (bps/Hz) &  $0.92, 0.85$ & $0.93, 0.82$ & $1.0, 0.83$ &  $1.04, 0.82$ \\
         \hline
         $(\nu_p, \Delta f)$  &  $2,30$ & $12,30$ & $8,30$ & $12, 30$ \\
         (in kHz) & & & & \\
         \hline
    \end{tabular}}\normalsize
    \label{tab3}
\end{table}

\subsection{Low mobility and small cell scenario}
\label{subsec1}
We consider low mobility ($\nu_{max} < 1/(2T) = 500$ Hz) and small cells ($\tau_s = 1.15 \mu s < 1/B$). Table-\ref{tab3} lists optimal parameters as a function of increasing $\nu_{max}$. Since channel delay and Doppler spreads are small, CP-OFDM is able to choose a SCS that avoids both ICI and ISI. The coherence time and coherence bandwidth are high (see (\ref{eqn9})) so that it is possible to acquire the CP-OFDM I/O relation with small pilot overhead. \emph{Zak-OTFS is only slightly more spectrally efficient than CP-OFDM, and the difference is due to higher CP overhead in CP-OFDM (Table-\ref{tab3}).} 


\subsection{High mobility and small cells}
\label{subsec2}
We consider high mobility ($\nu_{max} > 1/(2T) = 500$ Hz) and small cells ($\tau_s = 1.15 \mu s < 1/B$). Channel delay spread is small, but because Doppler spread is high, CP-OFDM needs to choose a high SCS to avoid ICI and ISI. Coherence time is small, since Doppler spread is high. Even a high SCS like $60$ kHz allows CP-OFDM to avoid ISI completely (since delay spread is small) and enables more dense pilot allocation in time (due to smaller symbol duration $T = 1/\Delta f$). However, it results in coarse pilot allocation in frequency and less accurate estimation of the I/O relation. In Table-\ref{tab3}, this is why the optimal SCS for $\nu_{max} = 1600$ Hz is $30$ kHz rather than $60$ kHz. \emph{For higher Doppler shifts, Zak-OTFS is $20$-$25\%$ more spectrally efficient than CP-OFDM, and the difference is due to more accurate acquisition of the I/O relation.}  

   \begin{table}[!t]
   \vspace{-4mm}
    \centering
    \caption{Optimal param. vs $\nu_{max}$, fixed $\tau_s = 4.7 \mu s$. (Zak-OTFS, CP-OFDM) for each parameter.}
   \scriptsize{
    \begin{tabular}{|c|c|c|c|c|}
         \hline
         $\nu_{max}$ (in Hz) &  $100$ & $400$ & $800$ &  $1600$ \\
         \hline
         MCS &  $8,7$ & $8,7$ & $8,6$ &  $9, 3$ \\
         \hline
         Overhead ($\%$) & $3.7, 13.3$ & $12.0, 13.3$ & $10.2, 15.5$ &  $20.0, 15.5$ \\
         \hline
         BLER  & $0.1, 0.1$ & $0.1, 0.09$ & $0.09, 0.1$ &  $0.1, 0.08$\\
         \hline
         SE (bps/Hz)  & $1.02, 0.8$ & $0.93, 0.81$ & $0.96, 0.67$ &  $0.95, 0.38$ \\
         \hline
         $(\nu_p, \Delta f)$   & $2,15$ & $6,15$ & $6,15$ & $12, 15$ \\
         (in kHz) & & & & \\
         \hline
    \end{tabular}}\normalsize
    \label{tab4}
    \vspace{-3mm}
\end{table}
\subsection{Low mobility and large cells}
\label{subsec3}
We consider low mobility ($\nu_{max} < 1/(2T) = 500$ Hz) and large cells ($\tau_s = 4.7 \, \mu s > 1/B$). Table-\ref{tab4} lists optimal parameters as a function of increasing $\nu_{max}$. High channel delay spread results in small coherence bandwidth necessitating an increase in pilot density in the frequency domain. Since pilot density cannot be very high due to the associated overhead, CP-OFDM I/O acquisition is inaccurate. In contrast, we can acquire the Zak-OTFS I/O relation with low pilot/guard overhead $\approx 2 \tau_s/\tau_p$ because $\nu_p$ is small and $\tau_p$ is large as the channel Doppler spread is small (see the optimal parameters for $\nu_{max} = 100$ Hz in Table-\ref{tab4}). \emph{For low mobility and large cells, Zak-OTFS is $15-20\%$ more spectrally efficient than CP-OFDM, and the difference is due to more accurate acquisition of the I/O relation given a similar pilot/guard overhead (see also discussion in Section \ref{subsecpreventici}).}


\subsection{High mobility and large cells}
\label{subsec4}
We consider high mobility ($\nu_{max} > 1/(2T) = 500$ Hz) and large cells ($\tau_s = 4.7 \mu s > 1/B$). The SE of CP-OFDM is poor because it is not possible to choose a SCS that avoids both ICI and ISI. A higher SCS avoids ICI but results in ISI, and a smaller SCS avoids ISI but results in ISI. In contrast, Zak-OTFS SE is almost independent of the Doppler shift because, rather than avoiding ICI, Zak-OTFS embraces ICI and equalizes it. Zak-OTFS I/O relation can be acquired accurately and efficiently (see Sections\ref{subseciorel2} and \ref{subsecpreventici} and the optimal parameters for $\nu_{max} = 1600$ Hz in Table-\ref{tab4}). \emph{For high mobility and large cells, Zak-OTFS is more than $50\%$ more spectrally efficient than CP-OFDM, because it is possible to accurately and efficiently estimate the Zak-OTFS I/O relation and it is not possible to accurately estimate the CP-OFDM I/O relation.}

\subsection{Connecting channel spread with accurate and efficient acquisition of the I/O relation}
The time-frequency representation of a doubly spread channel is not stationary, and a higher delay/Doppler spread implies a smaller coherence bandwidth/coherence time. When the pilot overhead is limited, it is not possible to accurately estimate the CP-OFDM I/O relation. \emph{In CP-OFDM the accuracy and efficiency of estimating the I/O relation is closely coupled to the channel delay and Doppler spreads.}

With Zak-OTFS, it is possible to accurately estimate the I/O relation provided the crystallization conditions are satisfied. It is possible to choose a single value of the Doppler period $\nu_p$ to support accurate and efficient estimation of the I/O relation regardless of channel spreads. By choosing $\nu_p = 12$ kHz, $\tau_p = 1/ \nu_p = 83.3 \mu s$, corresponding to $M=56, N=12$, we satisfy the crystallization conditions for every scenario considered in this paper. The pilot/guard overhead might not be optimal, but it is universal and good for every scenario. Since the pilot region is a Doppler strip, the highest overhead corresponds to the maximum delay spread $\tau_s = 4.7 \mu s$ and is $(2k_{max}+3)/M = 11/56$ for pilot allocation IV, which is less than $20\%$.

\section{Conclusions}
We have compared the CP-OFDM and Zak-OTFS SE across a full range of 6G propagation environments, representing the choice of waveform as an architectural choice between avoiding interference (CP-OFDM) and embracing interference (Zak-OTFS). Our comparison illuminates how relative SE depends on mobility (high/low) and on cell size (large/small). We find that while Zak-OTFS and CP-OFDM achieve similar SE for low mobility small cell scenarios, Zak-OTFS shows significant gains when either the delay spread, or the Doppler spread is high. This is because in CP-OFDM there is a limit on the maximum possible channel estimation resource (DMRS), which limits accurate estimation of the I/O relation when either the coherence time or coherence bandwidth is too small.

\end{document}